\documentclass[]{aa}
\usepackage{graphicx,hyperref,txfonts}
\usepackage{caption}
\usepackage{subcaption}
\usepackage{makecell}
\usepackage{array}
\usepackage{color}
\usepackage[normalem]{ulem}
\newcommand{\com}[1]{{#1}}

\newcommand{\pix}{{\tt{PixeLens}}}
\newcommand{\lensmodel}{{\tt{Lensmodel}}}
\definecolor{magen}{rgb}{0.70,0.10,0.50}
\definecolor{dgreen}{rgb}{0.1,0.5,0.2}

\begin{document}

\title{Model-independent and model-based local lensing properties of B0128+437 from resolved quasar images}
\titlerunning{Local lens properties of B0128+437}
\author{Jenny Wagner\inst{1} \and Liliya L.~R. Williams\inst{2}}
\institute{
Universit\"at Heidelberg, Zentrum f\"ur Astronomie, Astron. Rechen-Institut, M\"onchhofstr. 12--14, 69120 Heidelberg, Germany \\
\email{j.wagner@uni-heidelberg.de}
\and
School of Physics and Astronomy, University of Minnesota, 116 Church Street, Minneapolis, MN 55455, USA \\
\email{llrw@umn.edu}
}
\date{Received XX; accepted XX}

\abstract{The galaxy-scale gravitational lens B0128+437 generates a quadrupole-image configuration of a background quasar that shows milli-arcsecond-scale subcomponents in the multiple images observed with VLBI. 
As this multiple-image configuration including the subcomponents has eluded a parametric lens-model characterisation so far, we determine local lens properties at the positions of the multiple images with our model-independent approach.
Using \pix, we also succeed in setting up a global free-form mass density reconstruction including all subcomponents as constraints.
We compare the model-independent local lens properties with those obtained by \pix\ and those obtained by the parametric modelling algorithm \lensmodel. 
A comparison of all three approaches and a model-free analysis based on the relative polar angles of the multiple images corroborate the hypothesis that elliptically symmetric models are too simplistic to characterise the asymmetric mass density distribution of this lenticular or late-type galaxy. 
Determining the local lens properties model-independently, the sparsity and the strong alignment of the subcomponents yield broad $1$-$\sigma$ confidence intervals ranging from 8\% to over 1000\% of the local lens property values. The lens model approaches yield comparably broad confidence intervals. Within these intervals, there is a high degree of agreement between the model-independent local lens properties of our approach based on the subcomponent positions and the local lens properties obtained by \pix. 
In addition, the model-independent approach efficiently determines local lens properties on the scale of the quasar subcomponents, which are computationally intensive to obtain by free-form model-based approaches. 
Relying on the quadrupole moment of each subcomponent, these small-scale local lens properties show \com{tighter} $1$-$\sigma$ confidence bounds by at least one order of magnitude on the average with a range of 9\% to 535\% of the of the local lens property values. As only 40\% of the small-scale subcomponent local lens properties overlap within the confidence bounds, mass density gradients on milli-arcsecond scales cannot be excluded. Hence, aiming at a global reconstruction of the deflecting mass density distribution, increasingly detailed observations require flexible free-form models that allow for density fluctuations on milli-arcsecond scale to replace parametric ones, especially for asymmetric lenses or lenses with localised inhomogeneities like B0128.

}
\maketitle

\section{Introduction}
\label{sec:introduction}

B0128+437 is a galaxy-scale gravitational lens that has been controversially discussed in the literature. 
It was discovered by \cite{bib:Phillips_2000} as a gravitational lensing configuration of four multiple images of a quasar. 
Three of these images show three subcomponents each in the radio band, \cite{bib:Norbury_2002}, \cite{bib:Biggs_2004}. 
So far, no lens model with a simple, smooth mass distribution has been found that can explain the positions of all four quasar images and their subcomponents. 
Using the data from observations described in \cite{bib:Biggs_2004}, we further investigate this problem with the model-independent approach as developed in \cite{bib:Wagner2} to determine the ratios of scaled mass densities (convergences) and the reduced shear components at the angular positions of the multiple images. 
Subsequently, we compare these values to the ratios of scaled mass densities and reduced shear components that the parametric lens model approach \lensmodel\  (\cite{bib:Keeton_2001}, \cite{bib:gravlens}) and the free-form lens model approach \pix\  (\cite{bib:Saha_2004}) predict. 
Compared to previous lens models, we take into account the current best-fit redshifts of the lens and the source. 

This comparison is analogous to the one carried out in \cite{bib:Wagner4} on the galaxy-cluster scale lens CL0024 and shows that the model-independent approach can be applied to gravitational lensing configurations of any size in the same way. 
For both lens scales, the ratios of convergences and the reduced shear components of the model-independent approach show a high degree of agreement to the same local lens properties obtained by lens modelling approaches.
In this work, we investigate in addition if the model-independent approach is able to further constrain the local lens properties on the level of the subcomponents in the multiple images of the quasar behind B0128. 
Lens reconstructions like \pix~describe non-smooth irregularly shaped mass density distributions on a pixelised grid. Therefore, such methods usually require computationally intensive pixelisations to determine these small-scale properties.
For lens reconstructions based on analytical models like \lensmodel, solving the lens equation as an optimisation problem may also require a highly resolved sampling grid to determine small-scale details of the lens.
Thus, a more efficient way to obtain small-scale local lens properties is highly desired for the increasing amount of data in upcoming surveys.

The paper is organised as follows: 
Section~\ref{sec:related_work} introduces the information about B0128 that has become available so far. 
It is mainly based on the works of \cite{bib:Phillips_2000}, \cite{bib:Norbury_2002}, \cite{bib:Biggs_2004}, and \cite{bib:Lagattuta_2010}.
Subsequently, we list all observational data that we employ to calculate the model-independent, local lens properties and that are used to constrain our lens models in Section~\ref{sec:data}. 
In Section~\ref{sec:mi_ansatz}, we briefly outline the model-independent algorithm, which is further detailed in \cite{bib:Wagner2} before we show the model-independent, local lens characterisation of B0128. 
Analogously, Sections~\ref{sec:lm_ansatz} and \ref{sec:mb_ansatz} describe the lens modelling based on the parametric code \lensmodel, and the free-form code \pix, respectively, before applying it to B0128. 
To avoid any potential confirmation bias due to exchanging the values to be compared at an early stage, the evaluation is blinded in the same way as the evaluation performed in \cite{bib:Wagner4}. This means that the values of the local lens properties are determined independently for the model-independent and the model-based approaches and only revealed for the comparison at the very end, after all modelling is finished. 
In Section~\ref{sec:angles_ansatz}, we apply the model-free comparison as established in \cite{bib:Wold_2012,bib:Wold_2015} and \cite{bib:Gomer_2018} to B0128 to investigate its degree of asymmetry with respect to lenses with double mirror symmetry based on the relative polar angles between the multiple image positions. 
In Section~\ref{sec:conclusion}, we compare all results obtained in Sections~\ref{sec:mi_ansatz}, \ref{sec:lm_ansatz}, \ref{sec:mb_ansatz}, and \ref{sec:angles_ansatz}. Finally, we conclude by assembling a consistent picture of the lensing configuration in B0128 and summarising the methodological results that can be deduced from the comparison of the model-independent and the model-based lens reconstructions.

\section{Related work on B0128}
\label{sec:related_work}

\begin{figure*}[ht!]
\centering
  \includegraphics[width=0.4\textwidth]{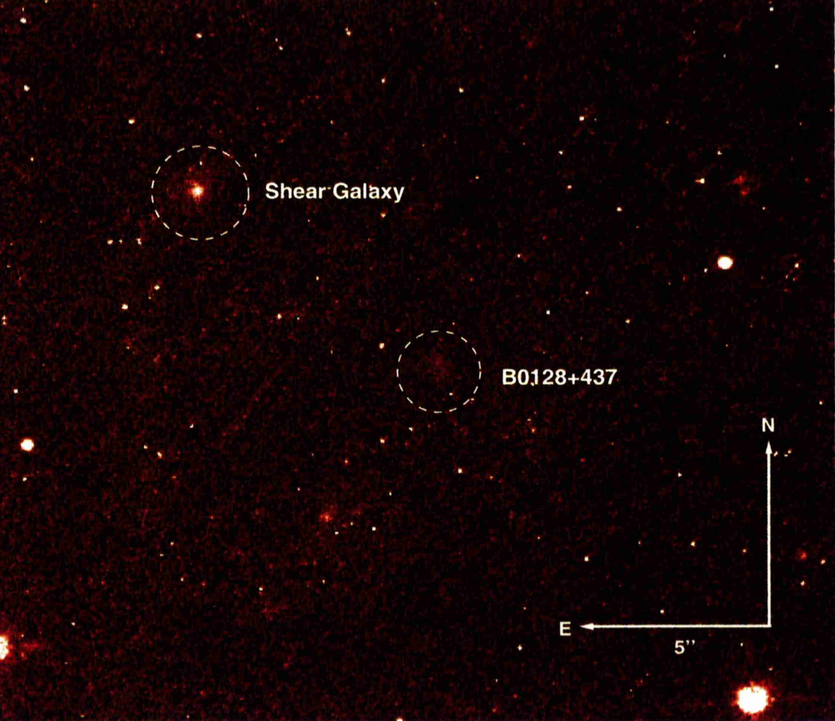} \hspace{0.1ex}
\includegraphics[width=0.4\textwidth]{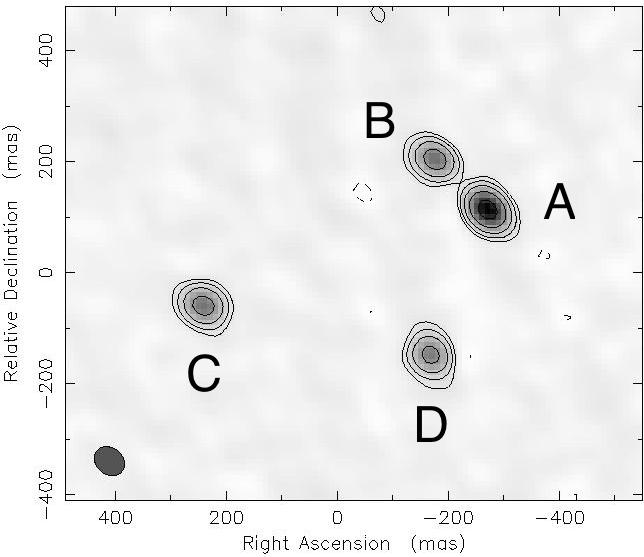} 
   \caption{Left: HST I-band observation from \cite{bib:Norbury_2002}; Right: MERLIN 5GHz observation by \cite{bib:Phillips_2000}.}
\label{fig:HST}
\end{figure*}

In the discovery paper, \cite{bib:Phillips_2000}, high-resolution MERLIN observations at 5 GHz from B0128 were obtained in the course of the Cosmic Lens All-Sky Survey (CLASS). 
Four unresolved multiple images, i.e. without visible substructures down to the scale of 30 mas, with a maximum image separation of 0.54 arcsec, arranged in a classic quad-lens formation were detected (see Figure~\ref{fig:HST} (right)). 
Long-term observations showed no hint of a time-variability of the quasar. 

B0128 was modelled by a singular isothermal ellipse (SIE), using the angular positions and flux ratios of the four (unresolved) multiple images to constrain the lens model. 
Redshifts of $z_\mathrm{l}=0.5$ for the lens and $z_\mathrm{s}=1.5$ for the source were assumed in a flat universe with $\Omega_{m0}=1$ as today's matter density parameter. 
High deviations between the observed and the model-predicted image positions were found. Including external shear alleviated the discrepancies, which indicated that the deflecting mass distribution might be less smooth than previously assumed. 

To investigate this phenomenon further, optical follow-up observations were performed as detailed in \cite{bib:Norbury_2002} (see Figure~\ref{fig:HST} (left) for an HST observation showing B0128 and its environment, including a galaxy that could be responsible for the external shear that was introduced in the SIE-lens model in \cite{bib:Phillips_2000}). 
They revealed that all signals from B0128 are highly reddened, such that the source is most probably at a high redshift or the lens contains a large amount of dust. 
In addition, follow-up VLBA 5 GHz data were obtained, showing that three of the four images (images $A$, $C$, and $D$ in Figure~\ref{fig:HST} (left)) could be further decomposed into three subcomponents. 

Subsequently, \cite{bib:Biggs_2004} summarised the results found in \cite{bib:Norbury_2002} to conclude that image $B$ in Figure~\ref{fig:HST} (right) is most likely scatter-broadened and hence, it is difficult to decompose it into subcomponents. 
\begin{figure*}[ht!]
\centering
  \includegraphics[width=0.38\textwidth]{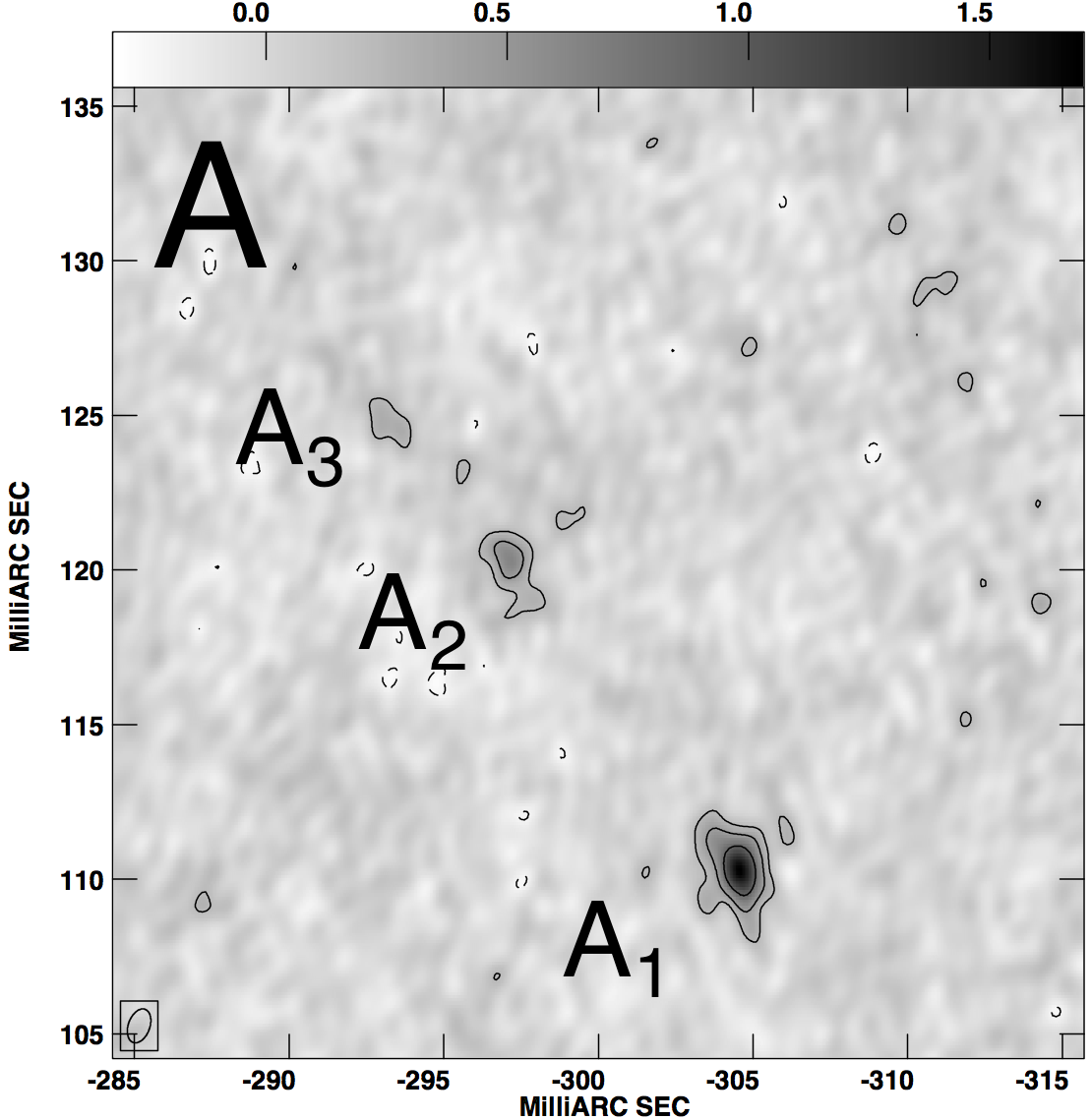} \hspace{4ex}
\includegraphics[width=0.38\textwidth]{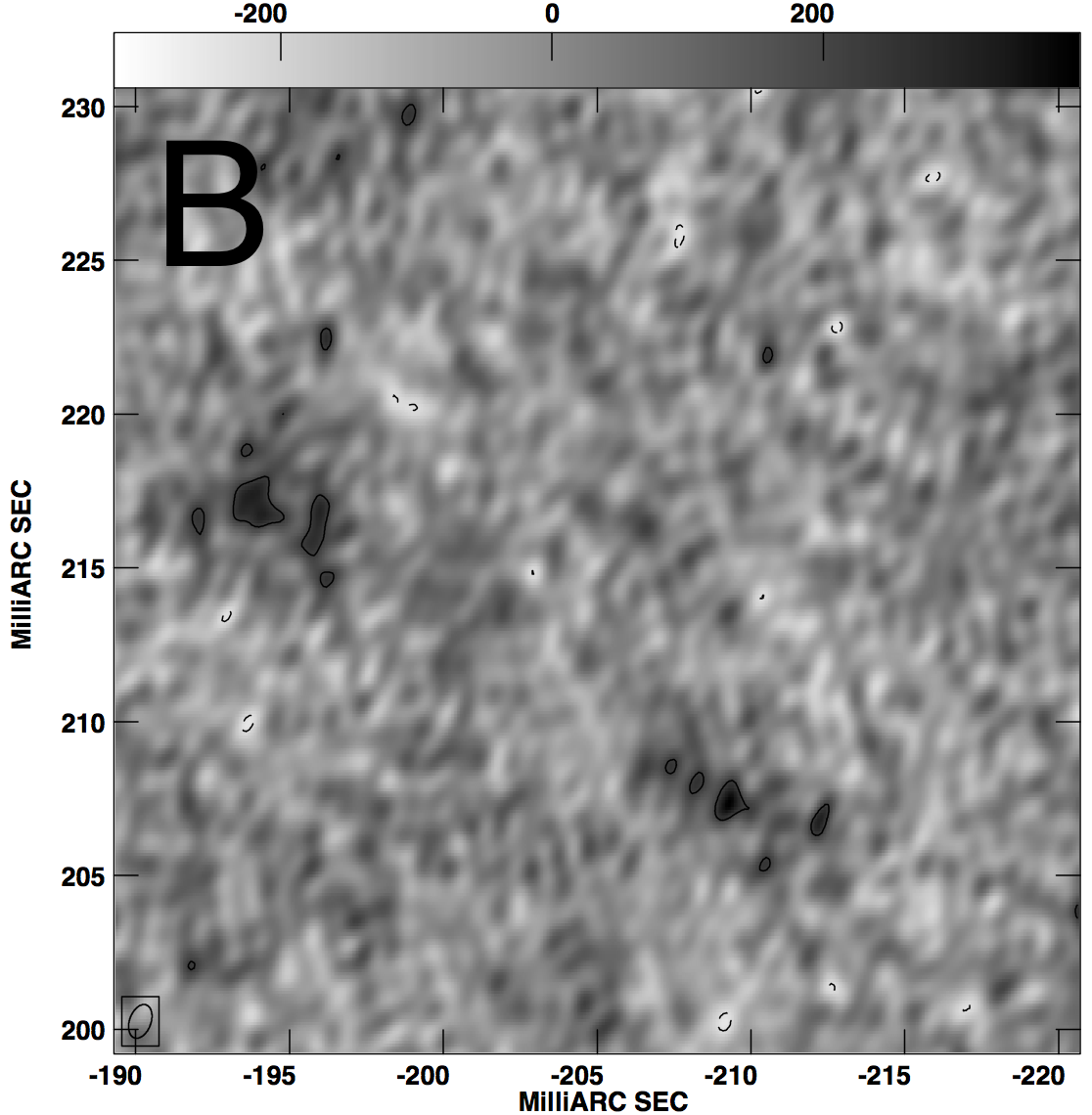} \\[4ex]
 \includegraphics[width=0.38\textwidth]{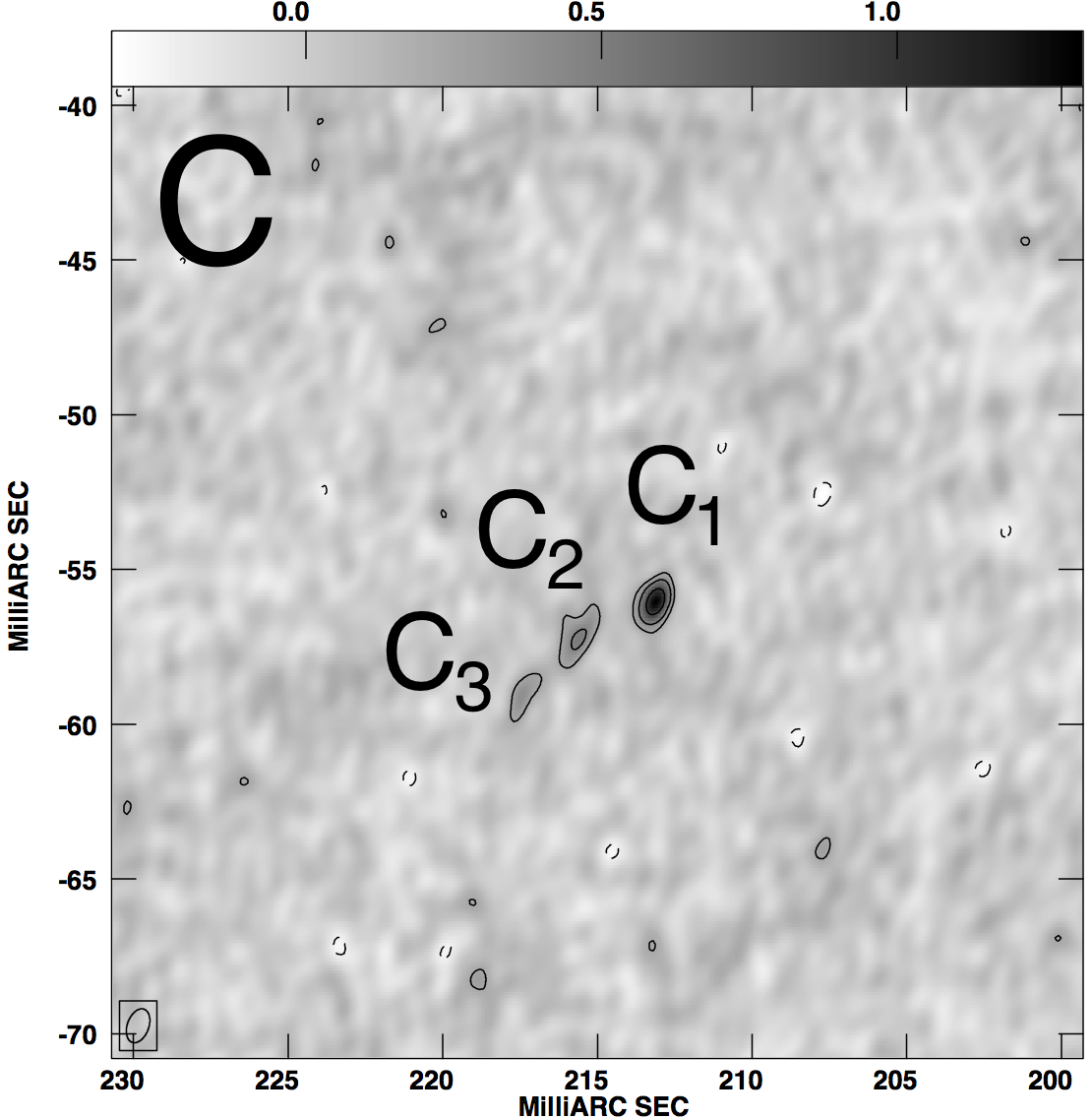} \hspace{4ex}
\includegraphics[width=0.38\textwidth]{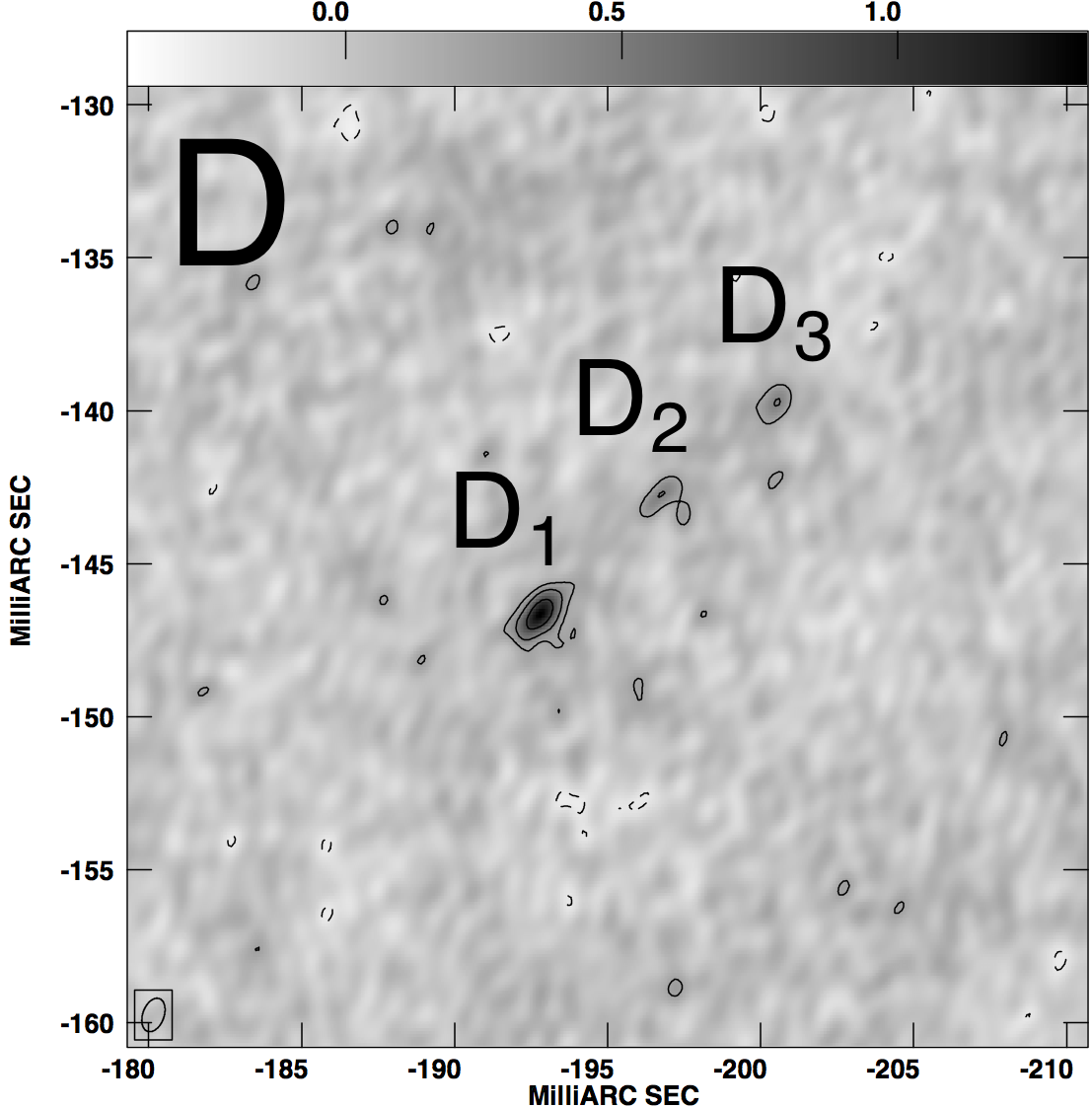} 
   \caption{VLBA 8.4 GHz details of all four multiple images from \cite{bib:Biggs_2004}.}
\label{fig:VLBA}
\end{figure*}
\begin{table*}[t]
 \caption{MERLIN 5 GHz measurements as in Table~2 of \cite{bib:Phillips_2000} (cols.~2--4), VLA 8.4 GHz measurements as in Table~1 of \cite{bib:Phillips_2000} (cols.~5--7), and as in Table~2 of \cite{bib:Lagattuta_2010} using the $Kp$-filter of the Near Infrared Camera 2 on the Keck II telescope along with the LGS AO system (cols.~8--10), both showing unresolved multiple images.}
\label{tab:unresolved}
\begin{center}
\begin{tabular}{cccccccccc}
\hline
\noalign{\smallskip}
Image & $F_{5} \left[ \mathrm{mJy} \right]$ & $\Delta \alpha_5 \left[ \mathrm{arcsec}\right]$ & $\Delta \delta_5 \left[ \mathrm{''}\right]$ & $F_{8} \left[ \mathrm{mJy} \right]$ & $\Delta \alpha_{8} \left[ \mathrm{''}\right]$ & $\Delta \delta_{8} \left[ \mathrm{''}\right]$ & $m_{K} \left[ \mathrm{mag} \right]$ & $\Delta \alpha_{K} \left[ \mathrm{''}\right]$ & $\Delta \delta_{K} \left[ \mathrm{''}\right]$\\
\noalign{\smallskip}
\hline
\noalign{\smallskip}
$A$ & $18.9$ & $\equiv 0.00$ & $\equiv 0.00$ & $14.8$ & $\equiv 0.00$ & $\equiv 0.00$ & $21.55 \pm 0.03$ & $\equiv 0.00$ & $\equiv 0.00$ \\
$B$ & $9.5$ & $0.098$ & $0.094$ & -- & -- & -- & $23.49 \pm 0.22$ & $0.099$ & $0.095$ \\
$C$ & $10.1$ & $0.520$ & $-0.172$ & $3.9$ & $0.497$ & $-0.188$ &$22.49 \pm 0.04$ & $0.521$ & $-0.170$ \\ 
$D$ & $9.2$ & $0.108$ & $-0.250$ & $5.8$ & $0.076$ & $-0.266$ & $22.87 \pm 0.12$ & $0.109$ & -0.260\\
\noalign{\smallskip}
\hline
\end{tabular}
\tablefoot{Col.~1: Name; Col.~2: Flux densities $F$ with their uncertainties of a few percent measured in the MERLIN 5 GHz band; Cols.~3 and 4: Relative angular image positions w.r.t. image $A$ at $\alpha = 01:31:13.405$, $\delta = +43^\circ 58' 13''.14$ (J2000.0); 
Col.~5: Flux densities $F$ with their uncertainties of a few percent measured in the VLA 8.4 GHz band, images $A$ and $B$ not separable; Cols.~6 and 7: Relative angular image positions w.r.t. image $A$ at $\alpha = 01:31:13.471$, $\delta = +43^\circ 58' 12''.938$ (J2000.0); 
Col.~8: Apparent magnitudes $m$ and their uncertainties as measured in the $Kp$-filter of NIRC2; Cols.~9 and 10: Relative angular image positions w.r.t. image $A$, reference position not given, uncertainties on images $B$, $C$, and $D$ are 0.001.}
\end{center}
\end{table*}
They also performed VLBI-observations at 2.3, 5, and 8.4 GHz to obtain further details about the subcomponents and found that the observations could rather show a core and jet instead of a compact symmetric object, as originally assumed in \cite{bib:Phillips_2000}. 
The detailed structure of all four images with the labelling of subcomponents according to \cite{bib:Norbury_2002} and \cite{bib:Biggs_2004} is shown in Figure~\ref{fig:VLBA}. 
Like the lens models by \cite{bib:Norbury_2002}, the lens models by \cite{bib:Biggs_2004}, based on the relative image positions as constraints, were not able to fully describe the lensing configuration to sub-milliarcsecond precision, even when different algorithms and modelling principles were employed. 

\cite{bib:McKean_2004} finally determined the redshift of the source to be $z_\mathrm{s} = 3.1240 \pm 0.0042$ in a KECK observation and found another emission line, which, depending on its origin, allows the lens to be at redshifts $z_\mathrm{l}=1.145, 0.645,$ or $0.218$. 
The assumption that the lens redshift is either $z_\mathrm{l}=0.645$ or $1.145$ was corroborated by further KECK observations as detailed in \cite{bib:Lagattuta_2010}. 
This work also corroborated all previous findings and set up the hypothesis that B0128 could be a lenticular or a late-type galaxy. 

Later, \cite{bib:Sluse_2012} modelled B0128 as an SIE with external shear as \cite{bib:Phillips_2000}, but used $\Omega_{m0} = 0.3$ and $\Omega_{\Lambda} = 0.7$ as today's cosmological parameters, in accordance with recent observations, requiring the cosmological model to introduce $\Omega_{\Lambda} \ne 0$. 
They assumed the lens at a redshift of $z_\mathrm{l} = 0.6$, which is less likely than $z_\mathrm{l}=1.145$, according to \cite{bib:Lagattuta_2010}. 
\cite{bib:Sluse_2012} conclude that the model fits the observation until sub-mas precision for the image positions is required. 
To investigate the most probable source of the discrepancies that enter at sub-mas precision, \cite{bib:Xu_2015} determined the probability that the flux anomalies between the images $A$ and $B$ are caused by yet unobserved dark matter substructures in the lens and concluded that the oversimplified or improper lens model in addition to the scatter-broadening is more likely to cause the flux anomaly than additional small-scale dark-matter halos.

Summarising the previous results, observations indicate that
\begin{itemize}
\item B0128 is a potentially lenticular or late-type galaxy most likely located at redshift $z_\mathrm{l}=1.145$ that generates four multiple images of a quasar at redshift $z_\mathrm{s}=3.124$,
\item three of these multiple images, $A$, $C$, and $D$ in Figure~\ref{fig:HST} (right), can be resolved into three clearly identifiable sub-components, the fourth, image $B$ in Figure~\ref{fig:HST} (right), is most likely scatter-broadened,
\item no smooth lens model (even including the shear galaxy located 7.8 arcseconds away, which is shown in Figure~\ref{fig:HST} (left)) can fully explain the positions of all sub-components to sub-mas-precision simultaneously. 
\end{itemize}

\section{Observed data from the multiple images}
\label{sec:data}

For our analysis, we use the data, as already stated in \cite{bib:Phillips_2000}, \cite{bib:Biggs_2004}, and \cite{bib:Lagattuta_2010}.
Tables~\ref{tab:unresolved} and \ref{tab:resolved} summarise the image positions and flux densities for the unresolved and resolved observations. Since it is very difficult to determine the relative angular positions for the hardly visible subcomponents in image $B$, we assume an uncertainty in these offsets of 1 mas and also investigate the impact, if we increase the uncertainty to 3 mas. 
A similar approach was pursued in \cite{bib:Biggs_2004}. 

In order to align the coordinates of the KECK observations in the optical from \cite{bib:Lagattuta_2010} with the MERLIN observations from \cite{bib:Phillips_2000}, we have to identify a reference point in both observations. 
To do so, we choose the angular position of image $C$, since the resolved components lie closest together and, as the counter image opposite to the three-image configuration $B$, $A$, $D$, it is farther from the critical curve as those images and is therefore subject to the least amount of magnification.

\begin{table*}[t]
 \caption{VLBI observations as in Table~1 from \cite{bib:Biggs_2004}.}
\label{tab:resolved}
\begin{center}
\begin{tabular}{crrrrrrr}
\hline
\noalign{\smallskip}
Image & $\nu \left[ \mathrm{GHz}\right]$ & $F \left[ \mathrm{mJy} \right]$ & $\Delta \alpha \left[ \mathrm{mas}\right]$ & $\Delta \delta \left[ \mathrm{mas}\right]$ & $a \left[ \mathrm{mas}\right]$ & $r$ & $\theta  \left[ \mathrm{deg}\right]$\\
\noalign{\smallskip}
\hline
\noalign{\smallskip}
$A_1$ & 5 & $3.9 \pm 0.1$ & $-304.3$ & $110.4$ & $4.2\pm0.2$ & $0.27\pm0.04$ & $27.0\pm2.1$ \\
$A_2$ & 5 & $3.1 \pm 0.2$ & $-297.1$ & $120.6$ & $2.5\pm0.2$ & $0.20\pm0.09$ & $28.0\pm2.9$\\
$A_3$ & 5 & $4.0 \pm 0.2$ & $-293.4$ & $124.4$ &$5.8\pm0.4$ & $0.16\pm0.03$ & $30.8\pm1.5$\\
\noalign{\smallskip}
\hline
\noalign{\smallskip}
$B_1$ & 5 & -- & $-197.0$ & $212.0$ & -- & -- & --\\
$B_2$ & 5 & -- & $-207.0$ & $210.0$ & -- & -- & --\\
$B_3$ & 5 & -- & $-210.0$ &  $207.9$ & -- & -- & --\\
\noalign{\smallskip}
\hline
\noalign{\smallskip}
$C_1$ & 5 & $2.7 \pm 0.1$ & $213.2$ & $-56.3$ & $3.1\pm0.2$ & $0.42\pm0.07$ & $-10.1\pm4.4$\\ 
$C_2$ & 5 & $1.2 \pm 0.1$ & $215.6$ & $-57.5$ & $1.9\pm0.4$ & -- & $-9.2\pm7.7$\\
$C_3$ & 5 & $1.6 \pm 0.1$ & $217.5$ & $-59.3$ & $2.0\pm0.3$ & $0.33\pm0.14$ & $-14.1\pm7.2$\\
\noalign{\smallskip}
\hline
\noalign{\smallskip}
$D_1$ & 5 & $2.4 \pm 0.1$ & $-193.0$ & $-146.7$ & $3.5\pm0.3$ & $0.34\pm0.06$ & $-46.6\pm3.9$ \\
$D_2$ & 5 & $2.0 \pm 0.1$ & $-196.6$ &  $-142.9$ & $1.7\pm0.2$ & $0.40\pm0.14$ & $-58.3\pm8.5$\\
$D_3$ & 5 & $1.4 \pm 0.1$ & $-200.1$ &  $-139.7$ & $2.0\pm0.3$ & $0.42\pm0.18$ & $-69.8\pm11.4$\\
\noalign{\smallskip}
\hline
\noalign{\smallskip}
$A_1$ & 8.4 & $3.6 \pm 0.3$ & $-304.5$ & $110.4$ & $1.6\pm0.2$ & $0.31\pm0.11$ & $29.0\pm6.5$ \\
$C_1$ & 8.4 & $1.3 \pm 0.2$ & $213.1$ & $-56.1$ & $0.4\pm0.3$ & --& $-35.1\pm37.3$\\ 
$D_1$ & 8.4 & $2.2 \pm 0.2$ & $-192.8$ & $-146.7$ & $1.1\pm0.2$ & $0.20\pm0.22$ & $-55.0\pm8.0$ \\
\noalign{\smallskip}
\hline
\end{tabular}
\tablefoot{Col.~1: Name of subcomponent; Col.~2: Observing frequency; Col.~3: Flux density and uncertainty; Cols.~4 and 5: Relative image positions to $\alpha = 01:31:13.494$, $\delta = +43^\circ 58' 12''. 805$ (J2000), extended by the components for image $B$ as read off Figure~6 in \cite{bib:Biggs_2004}. We assume an uncertainty of 0.1 mas for all subcomponent positions of images $A$, $C$, and $D$ and estimate an uncertainty of 1 mas for the subcomponent positions of image $B$; Col.~6: Extension of semi-major axis of elliptical Gaussian fitted by \textsc{Omfit}; Col.~7: Axis ratio of semi-minor to semi-major axis; Col.~8: Position angle measured from north to east.}
\end{center}
\end{table*}

The spectroscopic observation of \cite{bib:McKean_2004} did not find any hints of a second source and the similarity of the subcomponents in the images $A$, $C$, and $D$ is high, so that we treat the images $A$, $B$, $C$, and $D$ as multiple images from the same source at $z_\mathrm{s}=3.124$. 
We assume that the images $B$, $A$, and $D$ form a cusp configuration and that image $C$ is of the same parity as image $A$, see \cite{bib:Wagner2} for further details about fold and cusp configurations and their parity. 

Mapping the subcomponents of the images onto each other should then work analogously to the mapping of reference points in CL0024, as done in \cite{bib:Wagner4}. 
The subcomponents in the images in B0128 are much more aligned, which will cause higher uncertainties. 
In addition, three subcomponents are the minimum number of reference points required in each image to apply the method outlined in \cite{bib:Wagner2} and \cite{bib:Wagner4}. 
Considering images $B$ and $A$ as a fold configuration and images $A$ and $D$ as another fold configuration of two images mirror-inverted at the critical curve between them, we notice that the matching of subcomponents between $A$ and $D$ as shown in Figure~\ref{fig:VLBA} does not yield a fold configuration on the scale of the subcomponents. 
Matching the subcomponents by their intensity led to this labelling. 
Yet, the multi-band observations carried out in \cite{bib:Norbury_2002} and \cite{bib:Biggs_2004} strongly hint at dust in the lens, so that image $B$ might not be the only image that is subject to attentuation due to scatter-broadening. 
As a consequence, the matching of the subcomponents of image $A$ and $D$ could be different than proposed in \cite{bib:Biggs_2004}.
In Section~\ref{sec:mi_ansatz}, we systematically investigate the impact of different matchings on the local model-independent properties and the lens model. 

Table~\ref{tab:flux_ratios} shows the flux density ratios of images $B$, $C$, and $D$ with respect to image $A$ for all available bands, $\mathcal{F}_i$, $i=B, C, D$. \cite{bib:Biggs_2004} assume less than 5\% measurement uncertainties on their flux ratios. 
\cite{bib:Lagattuta_2010} obtain $\Delta \mathcal{F}_B = 3\%$, $\Delta \mathcal{F}_C = 1\%$, and $\Delta \mathcal{F}_D = 3\%$.
These flux density ratios are compared to the magnification ratios determined by the model-independent approach and the lens model as a consistency check because we do not employ them as constraints.

Natural weighting yields a high signal to noise ratio with a low angular resolution, such that it focuses on smooth, extended structures. 
Comparing the first two rows in Table~\ref{tab:flux_ratios}, we observe that the flux ratios are higher for this weighting scheme than for the uniform weighting which has a lower signal to noise ratio but a higher angular resolution. 
Hence, images $B$, $C$, and $D$ must be better visible, i.e. have a higher (flux) density, when the natural weighting scheme is applied. Furthermore, the fact that $\mathcal{F}_B$ is much smaller for the uniform than for the natural weighting scheme, strongly hints at scatter-broadening due to the dust in the lens, as already noted by \cite{bib:Biggs_2004}.


 \begin{table}[t]
 \caption{Observed flux density ratios $\mathcal{F}_i$ with respect to image $A$ for images $i=B, C, D$ in different bands.}
\label{tab:flux_ratios}
\begin{center}
\begin{tabular}{lccc}
\hline
\noalign{\smallskip}
Band & $ \mathcal{F}_B$ & $\mathcal{F}_C$ & $\mathcal{F}_D$ \\
\noalign{\smallskip}
\hline
\noalign{\smallskip}
VLBA 5~GHz (uni) & 0.26 & 0.45$^{(1)}$ & 0.45$^{(2)}$ \\
VLBA 5~GHz (nat) & 0.65 & 0.56 & 0.59 \\
VLBA 2.3~GHz & 0.49 & 0.34 & 0.47 \\
MERLIN 5~GHz & 0.56 & 0.49 & 0.47\\
$Kp$ & 0.17 & 0.42 & 0.30 \\
\noalign{\smallskip}
\hline
\end{tabular}
\tablefoot{Rows~1 and 2: In the 5-GHz VLBA maps with uniform (uni) and natural (nat) weighting as in Table~3 of \cite{bib:Biggs_2004}; Row~3: In 2.3~GHz VLBA maps; Row~4: In 5~GHz MERLIN maps as taken from Table~4 of \cite{bib:Biggs_2004}; Row~5: In the $Kp$-band (fifth row) as taken from Table~3 of \cite{bib:Lagattuta_2010};
\\
$^{(1)}$ flux density ratios for the subcomponents are $\mathcal{F}_{C,1} = 0.69$, $\mathcal{F}_{C,2} = 0.39$, $\mathcal{F}_{C,3} = 0.40$; 
\\
$^{(2)}$ flux density ratios for the subcomponents are $\mathcal{F}_{D,1} = 0.61$, $\mathcal{F}_{D,2} = 0.64$, $\mathcal{F}_{D,3} = 0.35$.}
\end{center}
\end{table}

\section{Model-independent reconstruction}
\label{sec:mi_ansatz}

\subsection{The method}
\label{sec:mi_ansatz_method}

Effectively describing the gravitational lens as a two-dimensional mass distribution in a single lens plane at redshift $z_\mathrm{l}$, the standard gravitational lensing formalism (see e.g. \cite{bib:Petters} or \cite{bib:SEF} for an introduction) treats this projected deflecting mass distribution in terms of the convergence $\kappa(\boldsymbol{x})$. 
This is the two-dimensional mass density at the position $\boldsymbol{x} \in \mathbb{R}^2$ in the lens plane scaled by the critical density $\Sigma_{\mathrm{cr}}$ which is the sufficient mass density to generate multiple images, given a lens at $z_\mathrm{l}$ and a source at $z_\mathrm{s}$. 
While $\kappa(\boldsymbol{x})$ only enlarges or diminishes the source, the shear $\boldsymbol{\gamma}(\boldsymbol{x}) = (\gamma_1(\boldsymbol{x}), \gamma_2(\boldsymbol{x}))$ additionally distorts the images. 
Both $\kappa(\boldsymbol{x})$ and $\boldsymbol{\gamma}(\boldsymbol{x})$ can be determined as second-order derivatives from the two-dimensional gravitational deflection potential $\psi(\boldsymbol{x})$ as detailed in \cite{bib:Wagner2}. 

Usually, global lens reconstructions, as detailed in Sections~\ref{sec:lm_ansatz} and \ref{sec:mb_ansatz} are set up, using the observables from the multiple images as constraints to reconstruct the deflecting mass density distribution as a whole. 
As observables, relative image positions, the quadrupole moment of the images around their centre of light, the flux ratios, and the time delays, if available, are employed.
These lens models are subject to a lot of degeneracies, see \cite{bib:Wagner5} for a mathematical derivation and \cite{bib:Wagner6} for the physical explanation of all degeneracies arising in the general lensing formalism.
To avoid the model-based degeneracies, \cite{bib:Tessore_2017} derived the most general information which can be obtained from multiple images of a background source without assuming a specific gravitational lens model. 
He found that transforming the multiple images onto each other yields ratios of convergences
\begin{equation}
f_{ij} \equiv \dfrac{1-\kappa(\boldsymbol{x}_i)}{1-\kappa(\boldsymbol{x}_j)} \equiv \dfrac{1-\kappa_i}{1-\kappa_j} 
\label{eq:f}
\end{equation}
between all multiple images $i,j$ and reduced shears
\begin{equation}
\boldsymbol{g}(\boldsymbol{x}_i) \equiv \boldsymbol{g}_i \equiv (g_{i,1}, g_{i,2}) \equiv \dfrac{\boldsymbol{\gamma}_i}{1- \kappa_i}
\label{eq:g}
\end{equation} 
at the positions of the multiple images. 
These local lens properties in Equations~\eqref{eq:f} and \eqref{eq:g} are invariant under the mass-sheet transformation. Thus, as further elaborated in \cite{bib:Wagner7}, they represent the information about the lens that all lens models should have in common at leading order.
Derived from the $f_{ij}$s and $\boldsymbol{g}_i$s, the magnification ratios
\begin{equation}
\mathcal{J}_{ij} \equiv \dfrac{\mu_j}{\mu_i} = \dfrac{\det(A_i)}{\det(A_j)}
\label{eq:J}
\end{equation} 
for image pairs $i,j$ can be calculated because the magnification $\mu_i$ of an image $i$ is given as the inverse of the determinant of the distortion matrix
\begin{align}
A_i = (1-\kappa_i) \left( \begin{matrix} 1-g_{i,1} & -g_{i,2} \\ -g_{i,2} & 1 + g_{i,1} \end{matrix} \right) \;.
\label{eq:A}
\end{align} 
As minimum requirements, the brightness profiles of the multiple images must contain clearly identifiable substructures, e.g. at least three linearly independent reference points like star forming regions, that can be matched across all multiple images. 
Furthermore, at least three multiple images with these identifiable substructures are needed which are not aligned like in an axisymmetric deflection potential. 

Our C-implementation\footnote{available at \url{https://github.com/ntessore/imagemap}} employs the centroid of all reference points as the position of a multiple image, also called ``anchor point". 
Apart from this choice, any point in the convex hull of the reference points can be used as anchor point, since the approach assumes that the convergence and the shear are approximately constant over the extension of a multiple image.
The position of the anchor point of one image, the so-called ``reference image" remains fixed. 
Then, a system of equations is set up by linearly mapping the reference image onto all other multiple images. 
Solving the system of equations by a $\chi^2$-parameter estimation as detailed in \cite{bib:Wagner4} yields all $f_ {ij}$ and $\textbf{g}_{i}$ for all multiple images, and the most likely anchor point positions in all remaining images. 
\cite{bib:Wagner2} and \cite{bib:Wagner4} further detail the algorithmic implementation and the procedure to obtain confidence bounds on the local lens properties by sampling their covariance matrix close to the most probable values of the $f_{ij}$, $\boldsymbol{g}_i$, and the anchor points.

\subsection{Results for B0128}
\label{sec:mi_ansatz_results}

Applying the method as outlined in Section~\ref{sec:mi_ansatz_method} to the case of B0128, we first find that the four non-aligned multiple images with their three subcomponents fulfil the requirements. 
The subcomponents are almost linearly aligned, yet, not in a mathematically rigorous way to cause the optimisation problem which determines the local lens properties to be under-constrained and thus degenerate. But, from the systematic analyses performed in \cite{bib:Wagner4}, we expect broad confidence bounds for the local lens properties determined by the subcomponents due to their alignment and the small area that they span. 

We choose image $A$ as the reference image and set up transformations between all remaining images to image $A$ in order to determine the local lens properties. 
To investigate the impact of this choice, we also determine the local lens properties in the three-image configuration of images $A, C,$ and $D$ mapping the three identifiable subcomponents onto each other and using image $C$ or image $D$ as reference image. 
In all three cases, we obtain the same local lens properties, yet with different confidence bounds. 
The analysis shows that using image $C$ as reference image yields smaller confidence bounds for the three-image configuration $ACD$. Yet, it yields comparable to slightly worse confidence bounds when matching the individual Gaussian fits of the subcomponents in image $A$, $C$, and $D$ onto each other and we obtain larger confidence bounds for the four-image configuration $ABCD$. 
Thus, image $C$ is only slightly less suitable as reference image than image $A$. 
Contrary to that, using image $D$ as reference image yields smaller confidence bounds for the three-image configuration, yet with a very low effective number of samples in the importance sampling process. 

To simplify the notation, we will omit $A$ in the subscripts of the local lens properties and simply write
\begin{equation}
f_{j} \equiv f_{Aj} = \dfrac{1 -\kappa_A}{ 1-\kappa_j} \;, \quad  \mathcal{J}_j \equiv \mathcal{J}_{Aj} = \dfrac{\mu_j}{ \mu_A} \;, \quad  j=B, C, D \;.
\label{eq:f_J_simplified}
\end{equation}

\begin{figure}[t!]
\centering
  \includegraphics[width=0.33\textwidth]{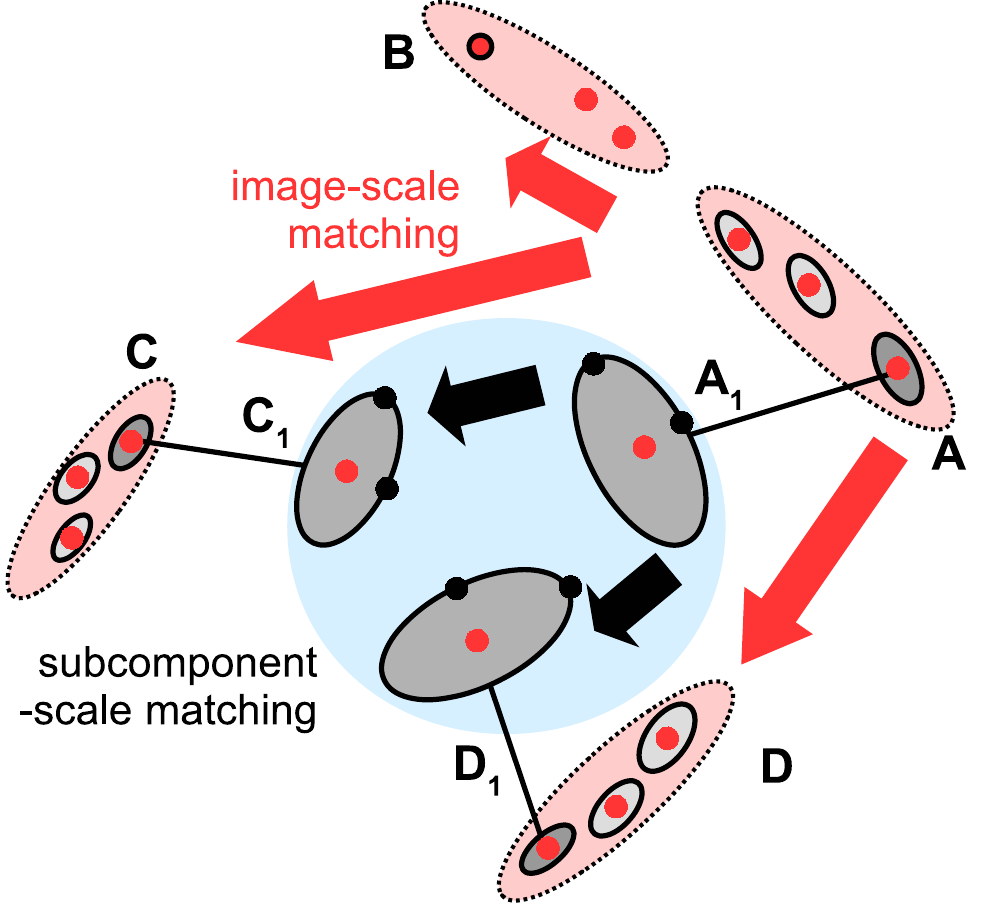}
\caption{Visualisation of image-scale matching: using the positions of the three subcomponents (red dots) in each image, the multiple images $A$, $B$, $C$, and $D$ can be matched onto each other (red arrows) to determine the local lens properties on the image scale (indicated by the larger red areas for visualisation purposes, as the convex hull spanned by the reference points in each image is too small to be drawn). 
Visualisation of subcomponent-scale matching (highlighted in blue): using the Gaussians fitted to a subcomponent (here: $1$, highlighted in dark grey), the subcomponents can be matched onto each other to determine the local lens properties within the area of the Gaussians. 
The first reference point in each Gaussian is its centre of light (red dots). 
The two reference points that are further required are the end points of the axes of the elliptical isocontour (black dots). 
North is to the top and east to the left.}
\label{fig:mi_matching}
\end{figure}

Instead of using the positions of the three subcomponents in each multiple image as reference points, we also reconstruct local lens properties on the scale of the subcomponents themselves. 
For this, we choose image $A$ again as reference image and use  the centre of light of the Gaussian fitted to the subcomponent $i$ as first reference point. 
Then, we use the end points of the semi-major and semi-minor axis of the fitted Gaussian as further reference points in each subcomponent. 
The calculations how to obtain these positions from the values of the semi-major axis, the axis ratio, and the position angle can be found in Appendix~\ref{app:Gauss_fits}. 
In this way, we can match the subcomponents $1$ and $3$ across images $A$, $C$, and $D$ to infer the individual local lens properties at their positions.
These values can be compared to the ones obtained for the convex hull of all subcomponents. 
The matching for the image- and subcomponent-scale is sketched in Fig.~\ref{fig:mi_matching}. 
%

In the following we start with matching the subcomponents for images $A$, $C$, and $D$ (see Section~\ref{sec:mi_ACD}). 
To investigate the  impact of a relabelling, we systematically interchange the $1$ and $3$ labels for images $C$ and $D$. 
Then, in Section~\ref{sec:mi_ABCD}, we include image $B$ and systematically interchange its subcomponents $1$ and $3$ to determine the most likely local lens properties for all four images. 
In Section~\ref{sec:mi_subcomponents}, we match the Gaussians fitted to subcomponents $1$ and $3$ across the images $A, C,$ and $D$ to investigate the local lens properties on the subcomponent-scale and discuss potential biases due to dust in the lens plane, as assumed in \cite{bib:Norbury_2002}, \cite{bib:Biggs_2004}, and \cite{bib:Lagattuta_2010}. 

\subsubsection{Image-scale matching of images $A$, $C$, and $D$}
\label{sec:mi_ACD}

Employing the positions of the three subcomponents $1$, $2$, and $3$ in images $A$, $C$, and $D$ listed in Table~\ref{tab:resolved} as reference points, we determine the local lens properties. 
The most likely lens properties, the mean values and the $1$-$\sigma$ confidence bounds are listed in the second, third, and fourth column of Table~\ref{tab:mi_results}, respectively. 

As mentioned in Section~\ref{sec:data}, the labelling of the subcomponents according to Table~\ref{tab:resolved} is not compatible with image $A$ and $D$ being a fold configuration on the level of subcomponents. 
To systematically investigate which subcomponents across the images should be matched, we interchange the labelling of the subcomponents $1$ and $3$ systematically as indicated in Table~\ref{tab:systematic_relabelling}. 
The resulting lens properties can be found in Appendix~\ref{app:mi_swapping_subcomponents}. None of the configurations obtains the correct relative parities between images $A$, $C$, and $D$ in the signs of the relative magnifications $\mathcal{J}_i$, $i=C, D$. 

\begin{table}[t]
 \caption{Configurations of differently matched subcomponents across images $A$, $C$, and $D$. Resulting local lens properties can be found in Table~\ref{tab:swapped_subcomponent_labellings}. Configuration 0 is the matching according to Table~\ref{tab:resolved}. In all  configurations, image $A$ is the reference image.}
\label{tab:systematic_relabelling}
\begin{center}
\begin{tabular}{cccc}
\hline
\noalign{\smallskip}
Conf. & Subcomp. 1  & Subcomp. 2 & Subcomp. 3 \\
\noalign{\smallskip}
\hline
\noalign{\smallskip}
   & $A_1$ & $A_2$ & $A_3$ \\
0 & $C_1$ & $C_2$ & $C_3$ \\
   & $D_1$ & $D_2$ & $D_3$ \\
\noalign{\smallskip}
\hline
\noalign{\smallskip}
   & $A_1$ & $A_2$ & $A_3$ \\
1 & $C_3$ & $C_2$ & $C_1$ \\
   & $D_1$ & $D_2$ & $D_3$ \\
\noalign{\smallskip}
\hline
\noalign{\smallskip}
   & $A_1$ & $A_2$ & $A_3$ \\
2 & $C_1$ & $C_2$ & $C_3$ \\
   & $D_3$ & $D_2$ & $D_1$ \\
\noalign{\smallskip}
\hline
\noalign{\smallskip}
   & $A_1$ & $A_2$ & $A_3$ \\
3 & $C_3$ & $C_2$ & $C_1$ \\
   & $D_3$ & $D_2$ & $D_1$ \\
\noalign{\smallskip}
\hline    
\end{tabular}
\end{center}
\end{table}

Thus, the labelling of subcomponents according to \cite{bib:Biggs_2004} is the only one yielding results which are consistent with leading order lensing theory. 
Considering the ratios of convergences $f_i$, we note that the highly negative value for $f_D$ could indicate that images $A$ and $D$ lie on opposite sides of the isocontour $\kappa(\boldsymbol{x}) = 1$. The positive sign for $f_C$ could indicate, that images $A$ and $C$ are located at the same side of $\kappa(\boldsymbol{x}) = 1$. 
Since $f_C$ and $f_D$ are both subject to broad confidence bounds several times their absolute value, these statements are further investigated in Sections~\ref{sec:mi_ABCD} and \ref{sec:mi_subcomponents}.
In Section~\ref{sec:mi_summary}, they are also checked for consistency in a comparison with our lens models as set up in Sections~\ref{sec:lm_ansatz} and \ref{sec:mb_ansatz}.

\subsubsection{Image-scale matching of all images}
\label{sec:mi_ABCD}

Building upon the configuration of three multiple images $A$, $C$, and $D$ with the three subcomponent positions as shown in Table~\ref{tab:resolved}, we include the three subcomponent positions in image $B$ and determine the local lens properties from it. 
The results can be read off columns~5--7 in Table~\ref{tab:mi_results} and are mostly subject to confidence bounds that exceed their absolute values. 
These large confidence bounds can partly be caused by the scatter-broadening of image $B$ which is not accounted for in our local lens reconstruction. 

Swapping the labelling of subcomponent $B_1$ with $B_3$, we find that both results agree within their confidence bounds except for $\boldsymbol{g}_{A}$. 
We favour the labelling as proposed by \cite{bib:Biggs_2004} due to the slightly lower confidence bounds of the resulting local lens properties and because the most likely value for $\mathcal{J}_B$ has the correct parity, which is not the case when we interchange the labels. 

In the four-image configuration, images $A$, $C$, and $D$ most likely lie on the same side of the isocontour $\kappa(\boldsymbol{x})=1$ and image $B$ seems to be located on the opposite side. 
Yet, analogously to the previous results, $f_B$ has a large confidence bound, which also includes the possibility to lie on the same side as image $A$.

Increasing the uncertainty in the positions of the subcomponents of image $B$ from 1 mas to 3 mas leads to an increase in the confidence bounds. 
The local lens properties still agree within their confidence bounds except for $\boldsymbol{g}_A$. 
Assuming an uncertainty of 3 mas in the position of all subcomponents, the confidence bounds do not increase, yet, the effective number of samples in the importance sampling is reduced below 10. 
In this case, the local lens properties except for $g_{A,1}$ coincide with the ones using 1 mas as uncertainty in the positions of the $B_i$, $i=1,2,3$. 
We show all results for the interchange of $B_1$ and $B_3$ and the increase of the measurement uncertainties in Appendix~\ref{app:mi_swapping_subcomponents2}. 

\begin{table*}[t]
 \caption{Synopsis of local lens properties obtained by the various configurations detailed in Sections~\ref{sec:mi_ACD}, \ref{sec:mi_ABCD}, and \ref{sec:mi_subcomponents}. $f_i$, $\boldsymbol{g}_i$ are listed with their most likely value, their mean and $1-\sigma$ confidence bound. ConfigACD is the image-scale three-image configuration of Section~\ref{sec:mi_ACD}, ConfigABCD is the image-scale four-image configuration of Section~\ref{sec:mi_ABCD}, ConfigA$_1$C$_1$D$_1$ is the subcomponent-scale three-image configuration at the position of subcomponent~$1$, ConfigA$_3$C$_3$D$_3$ is the subcomponent-scale three-image configuration at the position of  subcomponent~$3$. The \com{total} number of subcomponents (SCs) involved to determine the $f_i$ and $\boldsymbol{g}_i$ is listed below the name of the configuration.}
\label{tab:mi_results}
\begin{center}
\begin{tabular}{c|rrr|rrr|rrr|rrr}
\hline
\textbf{Config.} 	&		&   \textbf{ACD}	&		&		&	\textbf{ABCD}	&		&		&  \textbf{A$_1$C$_1$D$_1$}	&		&	&	\textbf{A$_3$C$_3$D$_3$}	&		\\
Lens prop.         &		&   \makecell[c]{9 SCs}	&		&		&	\makecell[c]{12 SCs}	&		&		&	\makecell[c]{3 SCs}	&		&	&	\makecell[c]{3 SCs}	&	\\
\hline																									
$\mathcal{J}_A$	&	1.00	&	1.00	&	0.00	&	1.00	&	1.00	&	0.00	&	1.00	&	1.00	&	0.00	&	1.00	&	1.00	&	0.00	\\
$f_A$	&	1.00	&	1.00	&	0.00	&	1.00	&	1.00	&	0.00	&	1.00	&	1.00	&	0.00	&	1.00	&	1.00	&	0.00	\\
$g_{A,1}$	&	-0.56	&	-0.56	&	0.04	&	-0.44	&	-0.44	&	0.05	&	3.53	&	3.62	&	0.73	&	1.32	&	1.40	&	0.39	\\
$g_{A,2}$	&	0.81	&	0.81	&	0.11	&	0.46	&	0.47	&	0.12	&	1.46	&	1.50	&	0.35	&	1.60	&	1.62	&	0.31	\\
\hline																									
$\mathcal{J}_B$	&		&		&		&	-1.47	&	-1.44	&	1.53	&		&		&		&		&		&		\\
$f_B$	&		&		&		&	-16.75	&	-0.64	&	122.09	&		&		&		&		&		&		\\
$g_{B,1}$	&		&		&		&	-8.77	&	0.31	&	66.85	&		&		&		&		&		&		\\
$g_{B,2}$	&		&		&		&	-6.09	&	-0.90	&	27.62	&		&		&		&		&		&		\\
\hline																									
$\mathcal{J}_C$	&	0.20	&	0.19	&	0.06	&	0.13	&	0.13	&	0.04	&	0.89	&	0.89	&	0.08	&	0.26	&	0.26	&	0.04	\\
$f_C$	&	1.44	&	1.86	&	35.99	&	0.29	&	0.34	&	0.82	&	0.67	&	0.68	&	0.13	&	0.45	&	0.46	&	0.11	\\
$g_{C,1}$	&	-0.54	&	-0.58	&	3.03	&	-0.16	&	-0.16	&	0.24	&	1.57	&	1.63	&	0.35	&	0.11	&	0.17	&	0.30	\\
$g_{C,2}$	&	0.55	&	1.21	&	45.83	&	-0.77	&	-0.71	&	1.07	&	2.31	&	2.36	&	0.43	&	1.88	&	1.90	&	0.43	\\
\hline																									
$\mathcal{J}_D$	&	-0.17	&	-0.17	&	0.10	&	-0.09	&	-0.10	&	0.09	&	-0.90	&	-0.90	&	0.11	&	-0.31	&	-0.31	&	0.06	\\
$f_D$	&	-6.92	&	-1.05	&	102.20	&	0.16	&	0.31	&	2.23	&	0.25	&	0.25	&	0.05	&	0.26	&	0.25	&	0.06	\\
$g_{D,1}$	&	-1.54	&	-0.08	&	29.03	&	0.15	&	0.21	&	0.55	&	0.21	&	0.20	&	0.09	&	-0.01	&	0.03	&	0.16	\\
$g_{D,2}$	&	3.07	&	-0.31	&	58.05	&	-1.07	&	-1.15	&	1.28	&	0.22	&	0.21	&	0.06	&	0.55	&	0.54	&	0.10	\\
\hline
\end{tabular}
\end{center}
\end{table*}

\subsubsection{Subcomponent-scale matching of subcomponents $1$ and $3$}
\label{sec:mi_subcomponents}

First, we investigate whether different configurations of reference points taken from the end points of the semi-major and semi-minor axes yield the same local lens properties. 
As detailed in Appendix~\ref{app:Gauss_fits}, this is the case for both subcomponents $1$ and $3$.
Subsequently, we determine the local lens  properties at the positions of subcomponent $1$ in images $A$, $C$, and $D$ as listed in columns~8--10 of Table~\ref{tab:mi_results} and the ones at the positions of subcomponent $3$ as listed in columns~11--13 of Table~\ref{tab:mi_results}.  
Contrary to the image-scale local lens properties, all $f$- and $\boldsymbol{g}$-values have confidence bounds that are smaller than their absolute value except for $g_{C,1}$ and $g_{D,1}$ of subcomponent 3. 
We find that half of the lens properties at the position of subcomponent $1$ agree with the ones at the position of subcomponent $3$ within their confidence bounds. 

Concerning their relative positions with respect to the isocontour $\kappa(\boldsymbol{x}) = 1$, the most likely $f_i$, $i=C, D$ inTable~\ref{tab:mi_results} imply that all images are supposed to lie on the same side to a much higher degree of confidence because the confidence bounds do not include negative values anymore.

\subsubsection{Synopsis and comparison to previous results}
\label{sec:mi_summary}

From Section~\ref{sec:mi_ACD}, we can draw the conclusion that it is possible to determine local $f_i$ and $\boldsymbol{g}_i$, $i=A, C, D$, such that their values are constant over the area spanned by the three subcomponents in each image. 
The matching of the subcomponents according to Table~\ref{tab:resolved} is the only one yielding the correct relative parities in the relative magnifications $\mathcal{J}_C$ and $\mathcal{J}_D$. 
These findings are in agreement with previous results as summarised in Section~\ref{sec:related_work}.

Including the subcomponent positions of image $B$, we find that the local lens properties for images $C$ and $D$ agree within their confidence bounds and all relative confidence bounds do not significantly increase. These results indicate that it should be feasible to construct a smooth lens model that explains the four-image configuration with its subcomponents. 
So far, only the specific approaches detailed in \cite{bib:Biggs_2004} have been unsuccessful.

In addition, we note that we obtained similarly large relative confidence bounds for the case of the galaxy-cluster-scale lens CL0024, see \cite{bib:Wagner4}, when we reduced the number of reference points to four that were almost aligned and spanned only a small area. 
Thus, apart from the scatter-broadening due to the comparably large amount of dust in the lens, the sparsity of the data and their alignment also contributes significantly to the large confidence bounds. 
The smaller confidence bounds determined for the local lens properties of the subcomponents due to the orthogonally oriented vectors between the reference points, support this hypothesis.
Without further, improved multi-band observations, it is hard to disentangle the impact of the individual effects.

Comparing the image-scale local lens properties as obtained in Section~\ref{sec:mi_ACD} to the local lens properties at the positions of the subcomponents $1$ and $3$, as derived in Section~\ref{sec:mi_subcomponents}, we find that the local lens properties at the positions of the two subcomponents agree with the ones determined over their entire convex hull with the exceptions of $\boldsymbol{g}_A$ at both positions and $\mathcal{J}_C$ and $\mathcal{J}_D$ at the position of subcomponent $1$.
Comparing the local lens properties at the positions of the two subcomponents with each other, we observe that half of them agree within their confidence bounds.
This puts a very weak upper limit to the scale of potential higher order perturbations, like gradients, in the surface mass density at the position of the images $A$, $C$, and $D$.

Local lens properties obtained at the positions of the subcomponents $1$ and $3$ indicate that all images lie on the same side of the isocontour $\kappa(\boldsymbol{x})=1$. 
Most probably, they lie outside $\kappa(\boldsymbol{x}) = 1$, given that $\kappa(\boldsymbol{x})= 1/2$ for an SIE and that such a smooth lens model can be fitted to the four-image configurations, if no sub-mas precision of the positions of the subcomponents is assumed.

\section{Model-based reconstruction: parametric}
\label{sec:lm_ansatz}

\subsection{The method}
\label{sec:lm_ansatz_method}

The most common way to model galaxy lenses is to fit image positions using a simple parametric form for the galaxy mass distribution. 
We use publicly available software, \lensmodel, described in \cite{bib:Keeton_2001} and \cite{bib:gravlens}. \lensmodel~offers a range of lens models; we choose to work with analytic potentials, instead of mass distributions, because the former have analytic expressions for all relevant lensing quantities. 
We note that the reconstructions presented in this section use a very different ansatz from the ones in Section~\ref{sec:mb_ansatz}, in that the latter recover a mass distribution, instead of a potential, and do not assume any fixed parameteric form.
The {\tt alphapot} lensing potential is a softened power law potential given by
\begin{equation}
\phi_\mathrm{a}(\boldsymbol{x})=b(s^2+\xi(\boldsymbol{x})^2)^{\alpha/2} \;, 
\end{equation}
where $b$ is the normalisation constant, $s$ is the core radius, which is set to a small non-zero value, and $\xi(\boldsymbol{x})^2=x_1^2+x_2^2/q^2$, with $q$ being the axis ratio of the potential. 
The boxy power law potential, called {\tt boxypot}, expressed in polar coordinates $\boldsymbol{x}=(r,\theta)$ has the form
\begin{equation}
\phi_\mathrm{b}(\boldsymbol{x})=br^{\alpha}[1-\epsilon\cos\,2(\theta-\theta_\epsilon)]^\beta \;,
\end{equation} 
where $\epsilon$ is the ellipticity and $\theta_\epsilon$ is its position angle\footnote{We note that \lensmodel~actually uses $\phi_\mathrm{b} = br^\alpha [1-\epsilon \cos 2(\theta - \theta_\epsilon)]^{\alpha\beta}$ to do the calculations, which is not exactly the same form as presented in the \cite{bib:gravlens} manual.}. 
In the reconstructions, both potentials are augmented with external shear.

\begin{figure*}[t]
\centering
\includegraphics[width=0.45\textwidth]{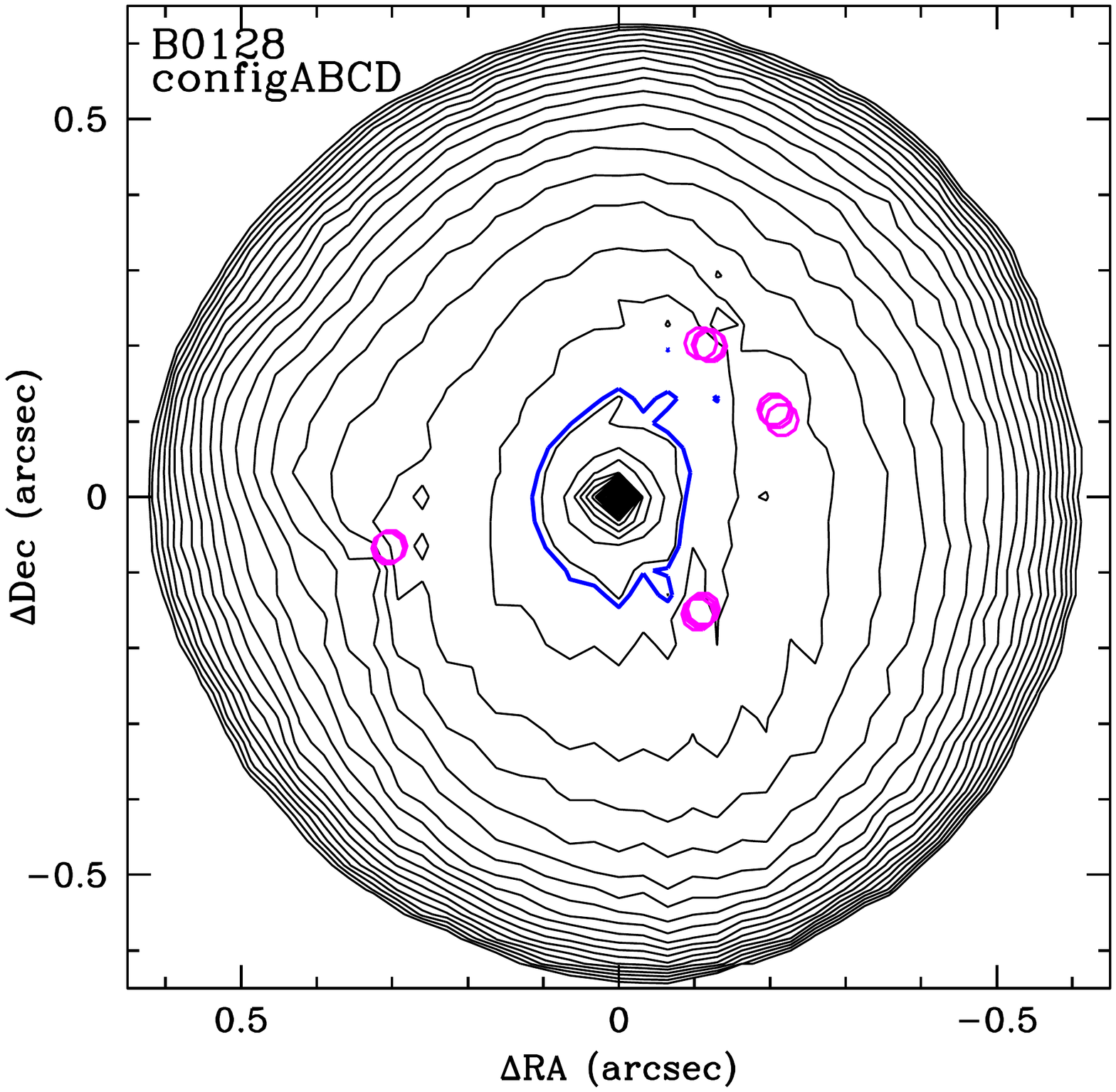} \hspace{0.1ex}
\includegraphics[width=0.45\textwidth]{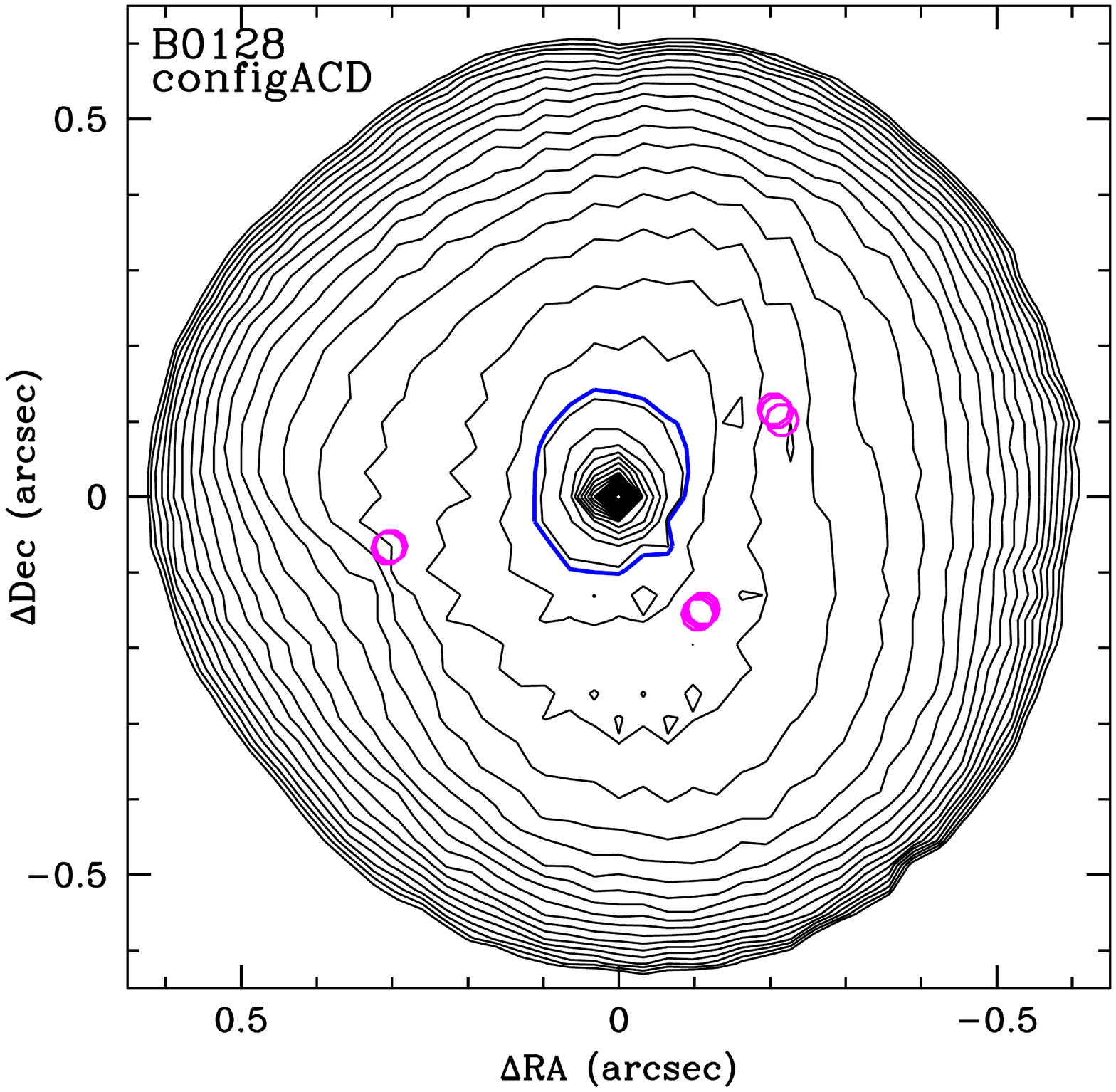} 
\vspace{-50pt}
   \caption{\pix~reconstructed maps of lensing convergence. Left: using all 12 subcomponents of B0128; Right: using only the 9 subcomponents of images $A$, $C$, and $D$. The subcomponents used as constraints in each case are indicated with magenta circles. The grey iso-convergence contours are spaced logarithmically. The blue contour is the iso-convergence contour at $\kappa(\boldsymbol{x})=1$.}
\label{fig:pixelensmass}
\end{figure*}
%

\subsection{Results for B0128}
\label{sec:lm_ansatz_results}

We run several lens reconstructions for each of the two image configurations (abbreviated as configABCD and configACD; we use the naming convention as introduced in Table~\ref{tab:mi_results}), and for the cases with and without fixing the galaxy lens centre. The reconstructions use different initial parameter guesses, and {\tt alphapot} and {\tt boxypot} potentials. 

If only subcomponent~1 is used as input, configA$_1$C$_1$D$_1$ and configA$_1$B$_1$C$_1$D$_1$ can be fit perfectly with either {\tt alphapot} or {\tt boxypot}, if the lens centre coordinates are left as free model parameters. 
This is summarised in the second column of the first two rows of Table~\ref{tab:lensmodel}, which shows typical root-mean-square deviations (rms) between the observed image positions and the model-predicted ones in the lens plane. 
If the lens centre is held fixed, configA$_1$C$_1$D$_1$ can still reproduce the images perfectly (third column of the first row). 
If configA$_1$B$_1$C$_1$D$_1$ is used, typical rms becomes $0.022\mathrm{''}$. 
When the other two subcomponents, 2 and 3, are included, for a total of 9 and 12 subcomponents for each of configACD and configABCD (last two rows), neither {\tt alphapot} nor {\tt boxypot} provide good fits, for floating or fixed lens centre. 
These lens plane rm are larger than our assumed uncertainty of 0.1 mas for images A, C, and D.


The resulting \lensmodel~local lens properties for configA$_1$B$_1$C$_1$D$_1$ and configA$_1$C$_1$D$_1$ as defined in Equations~\eqref{eq:f} and \eqref{eq:g} are extracted at the positions of subcomponent~1, and are listed in the columns 4--7 of Table~\ref{tab:evaluation_models}. 
The tightest confidence bounds are found for configA$_1$B$_1$C$_1$D$_1$ with fixed lens centre. This is probably because, in this case, the modelling has the least freedom to chose the lens mass distribution.

It is interesting to note that in B0128, ruling out simple lens models is possible because of excellent astromentric precision provided by the radio data (0.01 milli-arcsecond uncertainty), and further, by the presence of image substructure on scales smaller than $0.01\mathrm{''}$, i.e., substantially smaller than HST pixel size. If the source in B0128 were extended, with lensed images covering one or more HST-sized pixels in the lens plane, there would be no indication that simple models do not fit. 

 \begin{table}[h]
 \caption{\lensmodel~results: typical lens plane image rms. Naming convention according to the constraining observables as in Table~\ref{tab:mi_results}.}
\label{tab:lensmodel}
\begin{center}
\begin{tabular}{lccc}
\hline
\noalign{\smallskip}
Configuration & lens centre & lens centre \\
 & floating & fixed \\
\noalign{\smallskip}
\hline
\noalign{\smallskip}
configA$_1$C$_1$D$_1$ (3 SCs) & $0.0\mathrm{''}$ & $0.0\mathrm{''}$ \\
configA$_1$B$_1$C$_1$D$_1$ (4 SCs) & $0.0\mathrm{''}$ & $0.0022\mathrm{''}$ \\
configACD (9 SCs) & $0.0005\mathrm{''}$ & $0.0013\mathrm{''}$ \\
configABCD (12 SCs) & $0.0016\mathrm{''}$ & $0.0028\mathrm{''}$ \\
\noalign{\smallskip}
\hline
\end{tabular}
\end{center}
\end{table}

\section{Model-based reconstruction: free-form}
\label{sec:mb_ansatz}

\subsection{The method}
\label{sec:mb_ansatz_method}

Because simple parametric models, like elliptical mass distributions with external shear presented in Section~\ref{sec:lm_ansatz}, cannot reproduce all 12 subcomponents in B0128, here we use a free-form method, called \pix~\citep{bib:Saha_2004}, to reconstruct the mass density distribution in B0128. \pix~ is publicly available, and has an easy-to-use GUI interface\footnote{https://www.physik.uzh.ch/~psaha/lens/pixelens.php}.

For any given basis set, the lens equation can be written as a set of linear equations in the unknowns, which are the weights of the basis functions, and the source positions. \pix~breaks up the lens plane into equal size square mass pixels (its basis set), and imposes a few constraints. 
The positions of the images are specified with respect to the centre of the lens, which serves as the centre of the reconstruction. 
The parities of the images are also specified as input and are strictly enforced. 
The mass gradient should point not more than $\pm 45^\circ$ away from purely radial. 
Except for the central pixel, the mass of no other pixel can exceed twice the average mass of all its neighbours. 
These constraints act to regularise the mass distribution. 
Because many sets of pixel weights, i.e., many mass distributions, can reproduce the images exactly, there is a plethora of solutions. 
In this paper we run \pix~for 250 models, and discard the first 50 ``burn-in'' models, as is sometimes done for MCMC runs. 
The final  \pix~lens reconstruction is then taken as an average over 200 individual solutions. 

The confidence bounds for the local lens properties are calculated as the rms dispersion between 20 sets of 10 individual models. 
We use 10, instead of 1, because, averaging over a handful of individual mass density models suppresses one-off astrophysically unrealistic features and enhances common features. 
We experiment with averaging 10, 20 and 40 individual models to obtain one final \pix~model (with the corresponding number of 20, 10, and 5 sets of \pix~models to determine the rms).
As expected, we found that, the rms between the sets decreases. 
In the limit, if averaging is done over a very large number of individual reconstructions, the difference between sets will go to zero, and so will the rms. 
Our choice of averaging over 10 individual models to obtain one final model is somewhat conservative, because the corresponding rms is on the high side.

\subsection{Results for B0128}
\label{sec:mb_ansatz_results}

\begin{figure}[]
\centering
\includegraphics[width=0.45\textwidth]{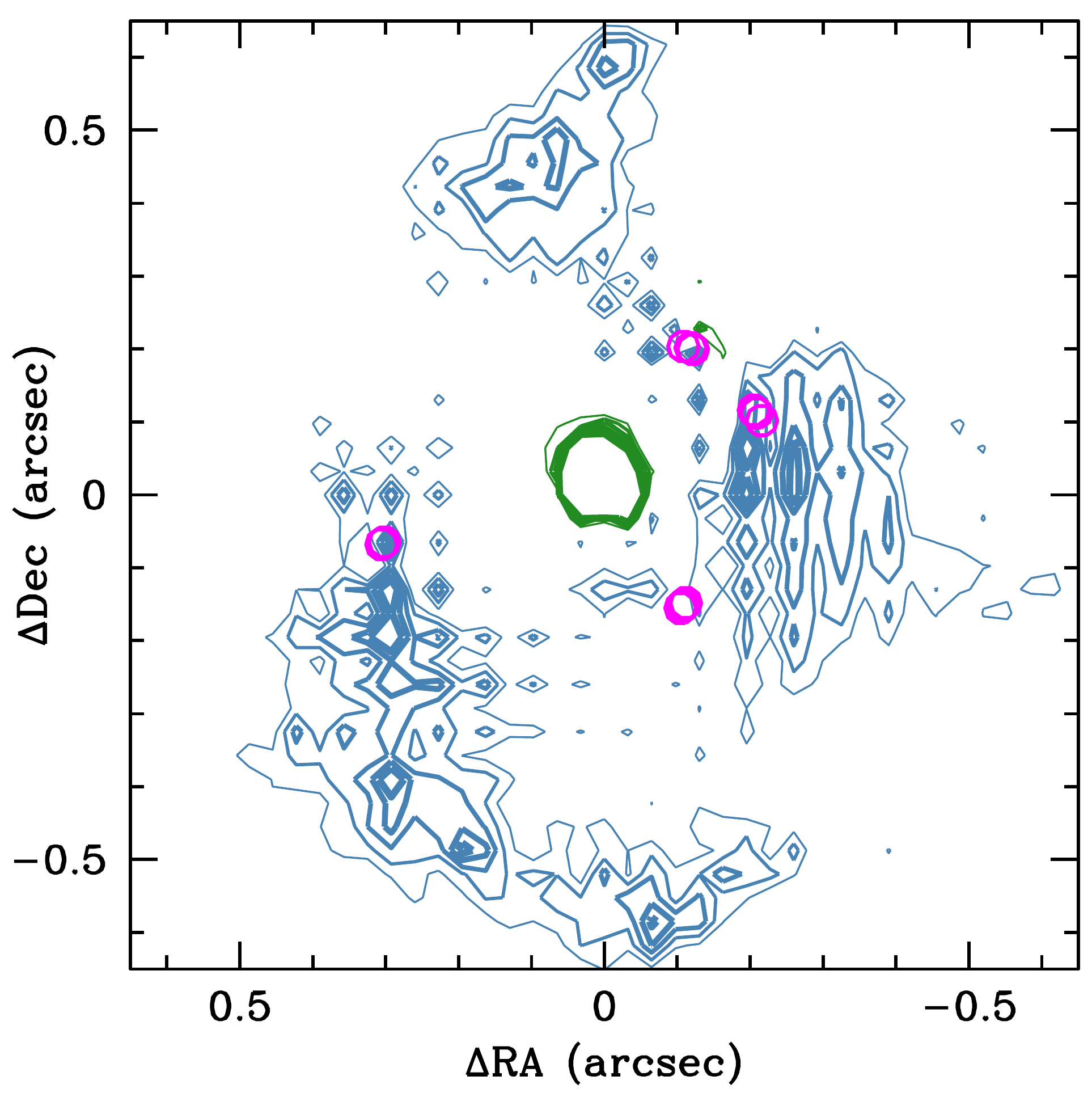}
   \caption{The result of subtracting mass density distribution of configABCD from that of configACD. Light blue (green) contours represent negative (positive) mass differences. Contour levels are at $\Delta\kappa$ values of $\pm 0.05,~\pm 0.065,~\pm 0.08, ~\pm 0.095, ~\pm 0.110$. None of the differences are statistically significant.}
\label{fig:pixelensdiff}
\end{figure}

First, we carry out \pix~reconstructions using all 12 subcomponents of all four images in B0128. Relative images fluxes are not used. 
A region of radius 0.675 arcseconds around the central reference point in Table~\ref{tab:resolved} is divided into 41 by 41 pixels, such that each \pix~pixel covers an area with an edge length of 33 milli-arcseconds.
The average projected lensing convergence map is shown in the left panel of Figure~\ref{fig:pixelensmass}. It is the average over 200 \pix~solutions.
The recovered mass distribution is not very circularly symmetric, and would be hard to represent with a simple parametric model. 
This is not surprising, and is consistent with parametric models not being able to reproduce all 12 subcomponents. 
Put differently, to reproduce all 12~subcomponents, one requires significant deviations from a purely elliptical projected mass distribution.

Mindful of the fact that mass reconstructions depend critically on the quality of the image data, we also carry out a reconstruction that does not include any subcomponents of the most-likely scatter-broadened image $B$. 
The reconstruction based on just the 3 subcomponents of each of images $A$, $C$, and $D$ is shown in the right panel of Figure~\ref{fig:pixelensmass}.
The local lens properties of configABCD and configACD as set up in Equations~\eqref{eq:f} and \eqref{eq:g} at the positions of subcomponent~1 are displayed in the second and third column of Table~\ref{tab:evaluation_models}.

Figure~\ref{fig:pixelensdiff} shows the differences between the convergence maps of the two models configACD and configABCD shown in Figure~\ref{fig:pixelensmass}. The light blue (green) contours represent negative (positive) differences between the convergences. To investigate whether these differences are signficant, we calculate the statistical significance at each location in the lens plane as
\begin{equation}
S=\dfrac{\kappa_A-\kappa_B}{\sqrt{\sigma_A^2+\sigma_B^2}} \;,
\label{eq:S}
\end{equation} 
where the $\sigma$'s are the rms obtained using 20 sets of 10 individual \pix~reconstructions. 
In fact, none of the differences shown in Figure~\ref{fig:pixelensdiff} are statistically significant; the value of $S$ is always less than 1.
Consistently, the $f$- and $\boldsymbol{g}$-values for the two models (see the first two models in Table~\ref{tab:evaluation_models}) all agree within their $1$-$\sigma$ confidence bounds.
Using alternative labelling of subcomponents, for example Conf.~1 in Table~\ref{tab:systematic_relabelling}, yields arrival time surfaces with very contorted contours, which is consistent with the results of Section~\ref{sec:mi_ACD}.

\begin{figure}[]
\centering
\includegraphics[width=0.45\textwidth]{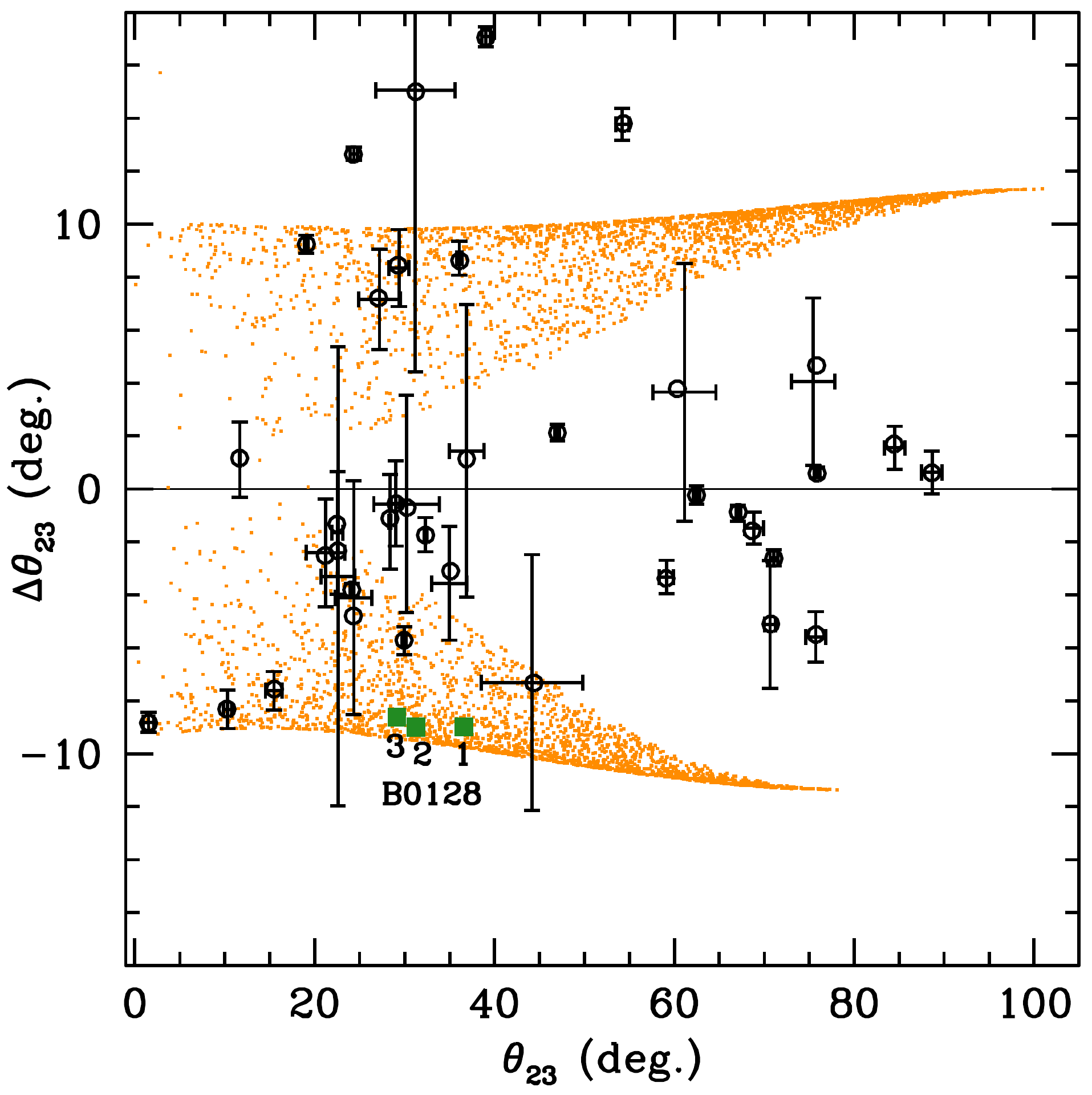}
   \caption{Distribution of quads in the space of relative image angles. Here, a 2D projection of that 3D space is used. The Fundamental Surface of Quads is the horizontal line at $\Delta\theta_{23}=0$. Galaxy-scale quads from \cite{bib:Wold_2012} are shown as circles with errorbars. The three quads of B0128 are represented by green squares, labeled with the corresponding subcomponent number. The orange points are quads from a synthetic lens, plotted here for reference, with projected density profile $\propto r^{-0.8}$, ellipticity $\epsilon=0.25$, and external shear $\gamma=0.25$, which is misaligned with the ellipticity position angle by $80^\circ$.  
}
\label{fig:angles}
\end{figure}
%

\begin{table*}
        \centering
        \caption{Local lens properties as defined by Equations~\eqref{eq:f} and \eqref{eq:g} of \pix~and \lensmodel~reconstructions for the different multiple-image configurations with subcomponents (details about the configurations, see Sections~\ref{sec:lm_ansatz} and \ref{sec:mb_ansatz}; naming conventions according to Table~\ref{tab:mi_results}).}
        \label{tab:evaluation_models}
        \begin{tabular}{c|r|r|r|r|r|r} 
                \hline

\textbf{Method}  & \makecell[c]{\textbf{\pix}} &  \makecell[c]{\textbf{\pix}} &  \makecell[c]{\textbf{\lensmodel}} &   \makecell[c]{\textbf{\lensmodel}} &  \makecell[c]{\textbf{\lensmodel}}  &   \makecell[c]{\textbf{\lensmodel}}  \\
\textbf{Config.}        &  \makecell[c]{\textbf{ABCD}}   &    \makecell[c]{\textbf{ACD}}    &    \makecell[c]{\textbf{A$_1$B$_1$C$_1$D$_1$}}   &     \makecell[c]{\textbf{A$_1$B$_1$C$_1$D$_1$}}   &   \makecell[c]{\textbf{A$_1$C$_1$D$_1$}}    &   \makecell[c]{\textbf{A$_1$C$_1$D$_1$}}    \\     
         &  \makecell[c]{12 SCs} &   \makecell[c]{9 SCs} &  \makecell[c]{3~SCs}  &    \makecell[c]{3~SCs}  &   \makecell[c]{3~ SCs}  &  \makecell[c]{3~SCs} \\
          &  \makecell[c]{centre fixed} & \makecell[c]{centre fixed} &  \makecell[c]{centre fixed} & \makecell[c]{centre floating} &  \makecell[c]{centre fixed} & \makecell[c]{centre floating} \\
Lens prop.   & & & &  & & \\
                \hline
 $\mathcal{J}_A$ & 1.00 $\pm$ 0.00 &   1.00 $\pm$ 0.00 &   1.0000 $\pm$ 0.0000 &   1.00 $\pm$ 0.00 &   1.00 $\pm$ 0.00 &   1.00 $\pm$ 0.00 \\  
$f_A   $          &  1.00 $\pm$ 0.00 &   1.00 $\pm$ 0.00 &   1.0000 $\pm$ 0.0000 &   1.00 $\pm$ 0.00 &   1.00 $\pm$ 0.00 &   1.00 $\pm$ 0.00 \\  
$g_{A,1}$      & -0.46 $\pm$ 0.16 &  -0.51 $\pm$ 0.13 &  -0.6158 $\pm$ 0.0026 &  -0.60 $\pm$ 0.11 &  -0.32 $\pm$ 0.20 &  -0.30 $\pm$ 0.56 \\  
$g_{A,2}$      & 0.60 $\pm$ 0.21 &   0.51 $\pm$ 0.13 &   0.3464 $\pm$ 0.0020 &   0.40 $\pm$ 0.11 &   0.41 $\pm$ 0.17 &   0.43 $\pm$ 0.45 \\  
                \hline
$\mathcal{J}_B$ & -0.26 $\pm$ 0.36 &   -0.29 $\pm$ 0.33 &  -0.6849 $\pm$ 0.0008 &  -0.69 $\pm$ 0.06 &   -0.60 $\pm$ 0.10 &   -0.52 $\pm$ 0.25 \\  
$f_B   $      &  3.80 $\pm$ 2.69 &   2.49 $\pm$ 0.89 &   1.6083 $\pm$ 0.0126 &   1.12 $\pm$ 2.54 &   5.43 $\pm$23.94 &  -1.00 $\pm$11.76 \\  
$g_{B,1}$      & 3.13 $\pm$ 3.88 &   1.79 $\pm$ 0.92 &   1.2594 $\pm$ 0.0011 &   0.88 $\pm$ 2.30 &   5.88 $\pm$26.30 &  -1.51 $\pm$13.84 \\  
$g_{B,2}$      & 2.44 $\pm$ 2.22 &  2.47 $\pm$ 1.20 &   1.1423 $\pm$ 0.0114 &   0.81 $\pm$ 1.63 &  -1.96 $\pm$13.80 &   1.79 $\pm$ 7.19 \\  
                \hline
$\mathcal{J}_C$ & 0.66 $\pm$ 2.26 &   0.76 $\pm$ 3.35 &   0.3589 $\pm$ 0.0007 &   0.36 $\pm$ 0.06 &   0.61 $\pm$ 0.19 &   0.66 $\pm$ 0.31 \\  
$f_C   $      & 0.76 $\pm$ 0.21 &   0.91 $\pm$ 0.18 &   0.7530 $\pm$ 0.0002 &   0.75 $\pm$ 0.13 &   0.87 $\pm$ 0.07 &   0.89 $\pm$ 0.14 \\  
$g_{C,1}$ &-0.63 $\pm$ 0.21 &  -0.78 $\pm$ 0.22 &  -0.4434 $\pm$ 0.0013 &  -0.45 $\pm$ 0.33 &  -0.29 $\pm$ 0.12 &  -0.24 $\pm$ 0.13 \\  
$g_{C,2}$ &-0.17 $\pm$ 0.08 &  -0.07 $\pm$ 0.05 &  -0.1108 $\pm$ 0.0004 &  -0.10 $\pm$ 0.08 &   0.13 $\pm$ 0.21 &   0.21 $\pm$ 0.26 \\  
                \hline
$\mathcal{J}_D$ & -0.16 $\pm$ 0.31 &  -0.21 $\pm$ 0.29 &   -0.4077 $\pm$ 0.0003 &   -0.39 $\pm$ 0.05 &   -0.40 $\pm$ 0.06 &   -0.35 $\pm$ 0.15 \\  
$f_D   $      & 1.56 $\pm$ 0.46 &   1.81 $\pm$ 1.11 &   1.0750 $\pm$ 0.0054 &   1.01 $\pm$ 1.16 &  -1.03 $\pm$13.85 &  -0.72 $\pm$ 7.15 \\  
$g_{D,1}$ & 1.31 $\pm$ 0.72 &  1.20 $\pm$ 0.48 &   0.4236 $\pm$ 0.0039 &   0.42 $\pm$ 0.87 &  -0.89 $\pm$ 9.23 &  -0.76 $\pm$ 4.91 \\  
$g_{D,2}$ &-2.08 $\pm$ 0.84 &  -2.13 $\pm$ 1.21 &  -1.4966 $\pm$ 0.0045 &  -1.42 $\pm$ 1.56 &   1.31 $\pm$16.63 &   1.11 $\pm$ 8.70 \\  
              \hline
        \end{tabular}
\end{table*}

\section{Model-free analysis}
\label{sec:angles_ansatz}

\subsection{The method}
\label{sec:angles_ansatz_method}

Finally, we perform another type of analysis on the 12 subcomponents of B0128, which is not a mass reconstruction, and so does not yield values of surface mass density or shear. 
The analysis is described in \cite{bib:Wold_2012,bib:Wold_2015} and \cite{bib:Gomer_2018}. 
It is based solely on the relative image polar angles of quadrupoly-imaged quasars (quads), as viewed from the centre of the galaxy lens. 
The three angles $\theta_{ij}$ are measured between the $i$th and $j$th arriving images: $\theta_{12}$, $\theta_{23}$, and $\theta_{34}$. 
\cite{bib:Wold_2012} show that all quads generated by lenses with double mirror symmetry, regardless of ellipticity or density profile slope, lie on a nearly invariant surface in the 3D space of the 3 image angles, called the Fundamental Surface of Quads (FSQ). 

Instead of using the 3D representation, it is easier to plot $\theta_{23}$ versus the deviation of quads from the FSQ, $\Delta\theta_{23}$. 
Note that $\theta_{23}$ is singled out because second and third arriving images are the ones that approach each other in the lens plane and vanish when the source moves further away from the lens centre and a quad becomes a double. 
Therefore these two images distinguish a quad from a double.

\subsection{Results for B0128}
\label{sec:angles_ansatz_results}

Figure~\ref{fig:angles} shows the three quads of the subcomponents in B0128 as green squares, together with 40 galaxy-scale quads presented in \cite{bib:Wold_2012}. In general, the more a given lens deviates from being purely double mirror symmetric, the more its quads will deviate from the FSQ. Deviations from double mirror symmetry can be of two general types: one can add external shear to an elliptical lens, or one can add non-elliptical mass density perturbations to an otherwise elliptical lens.

As a reference, we plot a few thousand quads from a synthetic lens with projected mass density profile $\propto r^{-0.8}$, ellipticity $\epsilon=0.25$, and external shear $\gamma=0.25$, which is misaligned with the ellipticity position angle by $80^\circ$. 
By comparing the location of the B0128 quads to the quads of the synthetic lens, we estimate that if B0128 is fitted with a simple lens model, its ellipticity and/or shear can be approximately $0.25$.
This is consistent with the findings of \cite{bib:Biggs_2004}, whose two models have $\gamma=0.26$ and $0.22$, as well as our own findings using \lensmodel~(Section~\ref{sec:lm_ansatz_results}), where ellipticity and external shear have similar magnitudes.

Since this is a rather large shear, an alternative interpretation of the location of the B0128 quads in Figure~\ref{fig:angles} is that the galaxy lens has non-elliptical density perturbations. 
The role of such perturbations on the relative image angles of galaxy-scale observed quads was explored by \cite{bib:Gomer_2018}. 
The authors concluded that observed deviations from FSQ by $\Delta\theta_{23}\sim 5^\circ\!-\!10^\circ$ are possible if realistic perturbations of the density profile from a purely elliptical model are included in the mass model. 

The possible presence of non-negligible perturbations from ellipticity in the case of B0128, indicated by Figure~\ref{fig:angles} is consistent with the mass distribution produced by free-form \pix, in Section~\ref{sec:mb_ansatz}, and shown in Figure~\ref{fig:pixelensmass}.


%
%

\section{Conclusion}
\label{sec:conclusion}

We present four different types of analyses and reconstructions of the galaxy-scale lenticular or late-type gravitational lens B0128 constrained by the quadrupole-image configuration of a background quasar.
Previous multi-band observations revealed that each of the four quasar images shows three bright subcomponent features. 
Radio observations resolve these subcomponents, which are separated by less than 10 milli-arcseconds, giving constraints on the lensing mass distribution on very small scales.
All approaches to find a global mass density reconstruction of B0128 based on lens models to reproduce the subcomponent-structure within the multiple images, mainly pursued by \cite{bib:Biggs_2004}, have been unsuccessful.
In contrast to that, the four multiple images observed at a lower resolution at which no subcomponents are resolved can be reproduced by a singular isothermal ellipse lens plus external shear.

Section~\ref{sec:methods} summarises the methodological progress that could be made to explain the multiple-image configuration at subcomponent-scale by analysing B0128 with different lens characterisation approaches and comparing their results with each other. 
Subsequently, we conclude in Section~\ref{sec:B0128} by summarising the consistent lens description that can be set up with all methods discussed and compared in Section~\ref{sec:methods}. 
Lastly, we put our findings in the context of similar cases.   

\subsection{Methodological results}
\label{sec:methods}

Using the model-independent approach, as detailed in \cite{bib:Wagner2} and \cite{bib:Wagner4}, we found lens-model-independent leading-order ratios of convergences and reduced shear values for all multiple images in B0128 based on the positions of the three subcomponents in the images (see Section~\ref{sec:mi_summary} and Table~\ref{tab:mi_results}). 
Subsequently, we succeeded in setting up a global free-form \pix~lens model using the positions of the subcomponents as constraints (see Section~\ref{sec:mb_ansatz_results}). 
Due to scatter-broadening and a strong alignment of the subcomponents, the local lens properties of all approaches are subject to broad confidence bounds. 
Within the $1$-$\sigma$ confidence bounds, the model-independent ratios of convergences and reduced shear values agree to the values obtained by the \pix~reconstruction in all but one.
So there is a similarly high degree of agreement between the model-independent local lens properties and the model-based values as was found in \cite{bib:Wagner4} for the galaxy-cluster-scale lens CL0024. 
Similar to \cite{bib:Wagner4}, we conclude that the overall width of confidence intervals of the local lens properties is decreased for the \pix~reconstruction with its additional global regularisation constraints compared to the model-independent ones.  
The tendency of tightening confidence intervals for an increasing amount of additional model assumptions and of regularisation constraints is supported by the findings of \cite{bib:Williams_2019}  on galaxy-cluster scale as well.

Determining the ratios of convergences and reduced shear values at the individual positions of the subcomponents, i.e.\@ on milli-arcsecond scale, we find larger mean reduced shear values than on image-scale.
In addition, the local lens properties between the subcomponents within one image only overlap in 50\% of the cases within their $1$-$\sigma$ confidence bounds.
The same degree of agreement is found when comparing the model-independent local lens properties at subcomponent~1 to the ones at the same positions obtained by the parametric \lensmodel~reconstruction using only subcomponent~1 in images $A$, $C$, and $D$ as multiple image constraints. 
The high amount of required milli-arcsecond-sized pixels in \pix~prevents us from setting up a \pix~model to determine the local lens properties at the individual subcomponents with their confidence bounds. 
Hence, \pix~is more robust than the model-independent approach in returning tighter confidence bounds due to additional regularisation constraints. Vice versa, the model-independent approach has the advantage over \pix~that it is highly efficient in returning local lens properties and their confidence bounds at any scale with a minimum amount of computational effort. 

On the whole, we conclude that the suitability of different global lens reconstruction approaches decisively depends on the resolution and the quality of the observations: 
on the scale of unresolved multiple images (e.g. for the data summarised in Table~\ref{tab:unresolved}), the mass density distribution in B0128 still has elliptical symmetry, so that parametric lens models like \lensmodel~are able to reproduce the multiple-image configuration, if the lens centre is not fixed.
The resolved subcomponent structures in the multiple images reveal asymmetries in the deflecting mass density distribution which require more sophisticated free-form modelling approaches like \pix~to explain the multiple-image configuration. 
Constraining local lens properties at the milli-arcsecond scale of the subcomponents to probe small-scale dark matter properties is computationally more efficient to pursue with the model-independent approach. 
It only yields local lens properties, i.e.\@ does not pursue a global reconstruction, but the local lens properties \com{give} the maximum information at leading order, which all lens model agree upon. 
Statistical screening methods that probe the symmetries of the observables of the multiple-image configuration like the one put forward in \cite{bib:Wold_2012} can serve as consistency checks or provide initialisations for the global lens reconstruction approaches to increase the modelling efficiency.

\subsection{Astrophysical conclusions for B0128}
\label{sec:B0128}

So far, only a few multiple-image configurations have been observed at milli-arcsecond level and have been found to show substructures on this scale. 
The quad-configuration in B0128 is one of these rare cases.
B0128 is also special in a second way because its deflecting mass density distribution is a high-redshift lenticular or late-type galaxy and not an early-type one.
As many recent works have consistently shown, see e.g.\@ \cite{bib:Hsueh_2018}, \cite{bib:Gomer_2018}, \cite{bib:Night_2019}, and \cite{bib:Gilman_2019}, smooth symmetric parametric lens models may not be a sufficient means to describe observed highly resolved multiple-image configurations on galaxy-scale much longer. 
The findings summarised in Section~\ref{sec:methods} consistently show that B0128 is such an example. 

Based on the \pix~model of Section~\ref{sec:mb_ansatz_results}, we confirm the hypothesis stated in \cite{bib:Xu_2015} that the lens models as set up in \cite{bib:Norbury_2002} or \cite{bib:Biggs_2004} are too simplistic to resolve the asymmetric deflecting mass density distribution that causes the multiple-images including their substructures on milli-arcsecond scale.
Given the type of the deflecting galaxy, it is not surprising that mass density isocontours change their morphology for increasing distance from the galactic centre. 
Taking into account the estimates of \cite{bib:Xu_2015} about potentially existing small-scale dark matter inhomogeneities in B0128 and the effects of the baryonic part of the mass density, the higher reduced shear values at the subcomponent-level and their potential variations between the subcomponents look plausible, but remain to be corroborated by further examples.

Comparing the magnification ratios obtained by the model-independent approach (see $\mathcal{J}_i$, $i=A, B, C, D$, in Table~\ref{tab:mi_results}), the magnification ratios as determined by \pix~(see $\mathcal{J}_i$ in Table~\ref{tab:evaluation_models}), and the observed flux ratios (see Table~\ref{tab:flux_ratios}), we find a high degree of agreement between the observed flux ratios and the \pix~values within the broad confidence bounds of the \pix~reconstruction.
The values based on the model-independent approach have tighter confidence bounds and only agree for the subcomponent 3 in image $D$ with the observed ones. 
While the comparison between the flux ratios and the \pix~magnification ratios is considered over areas that have the same order of magnitude, the model-independent approach determines the magnification ratios over the triangle spanned by the three subcomponent positions, which is less than 1\% of the area of a \pix~pixel.
For the subcomponents, the differences between observed flux ratios and model-independent magnification ratio values shrink, which corroborates the hypothesis that the different sizes over which the quantities are determined causes discrepancies between the results.
Yet, further investigations on the way that the flux ratios are calculated are necessary to confirm this. 
In addition, the observed flux ratios can be influenced by microlensing, scatter-broadening and absorption, see \cite{bib:Biggs_2004} and \cite{bib:Lagattuta_2010} for further details.

On the whole, we can conclude that, at the current observational accuracy and precision, we have arrived at a consistent reconstruction of the deflecting mass density distribution and model-independent local lens properties of B0128 which are able to explain the observed multiple-image configuration including the subcomponent structure on milli-arcsecond scale. The findings are also in accordance with observations and modelling results of previous works.
 
The high-precision astrometry and existence of subcomponents in the radio bands give an unprecedented view of the galaxy-scale lens. Evidence for deviations from a simple lens is found at arcsecond \emph{and} milli-arcsecond scales. On milli-arcsecond scale, there are high shear values and shear gradients, which imply gradients in mass density. On arcsecond scale, simple parametric mass distributions, like \lensmodel~\com{models} cannot reproduce images within the astrometric \com{precision (does that sound better? for me, it's strange to produce something within some error)} if the lens centre is fixed (see Section~\ref{sec:lm_ansatz_results}).  Furthermore, the external shear required (as determined by \lensmodel~and the model-free approach in Figure~\ref{fig:angles}) is 0.22-0.26. Such large shears are unlikely to arise from nearby galaxies. For comparison, \cite{bib:Bolton_2008} modelled 63 SLACS lenses and found that shears range from 0 to 0.27, but the median is only 0.05. The large shear value in B0128 could suggest that shear subsumes in it other complexities of the mass distribution, such as those suggested by \cite{bib:Gomer_2018}, and illustrateed in their Figure 14.

A similar case like B0128 is B1933+503, \cite{bib:Cohn_2001}, \cite{bib:Suyu_2012}, which is a spiral galaxy with 10 multiple images of a three-component source. 
Observations at the resolution of milli-arcsecond scale are not yet available.
A second, similar case is the galaxy-group of B1349+154, \cite{bib:Rusin_2001}, in which an unprecedented triangle of galaxies forms a gravitational lens that generates six multiple images of a background quasar. 
VLBA 1.7~GHz observations hint at potentially resolvable substructures in the brightest multiple images, so that this unique lensing configuration could also be an informative target.
Further candidates for future small-scale radio observations and analyses could be selected from the 40 quadrupole-image configurations in \cite{bib:Wold_2012} that show high distances to the Fundamental Surface of Quads.

\begin{acknowledgements}
The authors would like to thank Jori Liesenborgs for kindly agreeing to be the mediator of the unblinding process.
JW gratefully acknowledges the support by the Deutsche Forschungsgemeinschaft (DFG) WA3547/1-3.
\end{acknowledgements}

\bibliographystyle{aa}
\bibliography{aa}

\appendix

\section{Derivation of the reference point positions on the subcomponent-scale}
\label{app:Gauss_fits}

\subsection{Derivation of coordinate positions}
\label{app:reference_points}

\begin{figure}[h!]
\centering
  \includegraphics[width=0.23\textwidth]{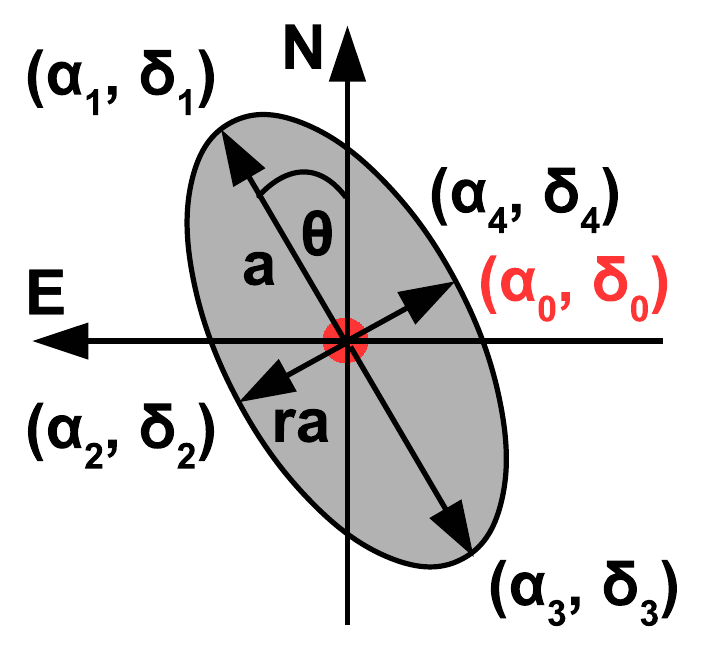}
\caption{Determining the reference points from an elliptical Gaussian fitted to the subcomponents: the centre of light is given by the relative coordinates $\Delta \alpha$ and $\Delta \delta$ in Table~\ref{tab:resolved}, here denoted by $(\alpha_0, \delta_0)$. 
For images $A$ and $C$ with the same parity, we use the positions of the semi-major and semi-minor axes as denoted by $(\alpha_1, \delta_1)$ and $(\alpha_2, \delta_2)$. Since image $D$ has opposite parity, we employ $(\alpha_1, \delta_1)$ and $(\alpha_3, \delta_3)$ as reference point positions in addition to $(\alpha_0, \delta_0)$.}
\label{fig:Gaussian_reference_points}
\end{figure}

\noindent
Fig.~\ref{fig:Gaussian_reference_points} depicts the quantities measured in the elliptical Gaussians fitted to the subcomponents: the length of the semi-major axis $a$, the axis ratio between the semi-minor and the semi-major axis $r$, and the position angle $\theta$ measured with respect to the north axis. 
Denoting the centre of light of the Gaussian at the relative coordinates $(\Delta \alpha, \Delta \delta)$ from a global reference point (see Table~\ref{tab:resolved}) as $(\alpha_0, \delta_0)$, we obtain the four end points at the axes of the Gaussian by the following trigonometric relations:
\begin{align}
\alpha_1 =& \, \alpha_0 + \phantom{r}a \sin(\theta) \;, &\delta_1 = \delta_0 + \phantom{r}a \cos(\theta) \;, \\
\alpha_2 =& \, \alpha_0 + ra \cos(\theta) \;, &\delta_2 = \delta_0 - ra \sin(\theta) \;, \\
\alpha_3 =& \, \alpha_0 - \phantom{r}a \sin(\theta) \;, &\delta_3 = \delta_0 - \phantom{r}a \cos(\theta) \;, \\
\alpha_4 =& \, \alpha_0 - ra \cos(\theta) \;,  &\delta_4 = \delta_0 + ra \sin(\theta) \;.
\end{align}

To estimate the uncertainty of the $(\alpha_i, \delta_i)$, $i=1,...,4$, we assume that the uncertainties in $\alpha_0, \delta_0, r,$ and $a$ were uncorrelated. 
Due to the fitting procedure, this is not the case. 
Yet, it yields the order of magnitude to which the reference points can be determined. Given the uncertainties in Table~\ref{tab:resolved}, we find uncertainties on the order of 0.1 to 0.2 mas for the subcomponents in Table~\ref{tab:resolved}.

\subsection{Impact of the uncertainties in the positions on the local lens properties}

When mapping one elliptically Gaussian subcomponent to the one of another multiple image, the mapping should be independent of the reference points used.
Only the relative parity between the multiple images must be obeyed. 
Thus, for the case of B0128, the configurations of reference points as listed in Table~\ref{tab:ref_point_selection} should all yield the same local lens properties.
Determining the reference points from the semi-major and semi-minor axes according to Appendix~\ref{app:reference_points} and calculating the local lens properties for all configurations listed in Table~\ref{tab:ref_point_selection}, we corrobate this assumption.

\begin{table}[t]
 \caption{Configurations of reference points in the subcomponents $i=1,3$ across images $A$, $C$, and $D$ that should all yield the same local lens properties.}
\label{tab:ref_point_selection}
\begin{center}
\begin{tabular}{cccc}
\hline
Conf. & $A_i$  & $C_i$ & $D_i$ \\
\hline
   & $(\alpha_0, \delta_0)$ & $(\alpha_0, \delta_0)$ & $(\alpha_0, \delta_0)$ \\
0 &$(\alpha_1, \delta_1)$ & $(\alpha_1, \delta_1)$ & $(\alpha_1, \delta_1)$ \\
   & $(\alpha_4, \delta_4)$ & $(\alpha_4, \delta_4)$ & $(\alpha_2, \delta_2)$ \\
\hline
   & $(\alpha_0, \delta_0)$ & $(\alpha_0, \delta_0)$ & $(\alpha_0, \delta_0)$ \\
1 &$(\alpha_1, \delta_1)$ & $(\alpha_1, \delta_1)$ & $(\alpha_1, \delta_1)$ \\
   & $(\alpha_2, \delta_2)$ & $(\alpha_2, \delta_2)$ & $(\alpha_4, \delta_4)$ \\ 
\hline
   & $(\alpha_0, \delta_0)$ & $(\alpha_0, \delta_0)$ & $(\alpha_0, \delta_0)$ \\
2 &$(\alpha_2, \delta_2)$ & $(\alpha_2, \delta_2)$ & $(\alpha_3, \delta_3)$ \\
   & $(\alpha_3, \delta_3)$ & $(\alpha_3, \delta_3)$ & $(\alpha_4, \delta_4)$ \\
\hline
   & $(\alpha_0, \delta_0)$ & $(\alpha_0, \delta_0)$ & $(\alpha_0, \delta_0)$ \\
3 &$(\alpha_3, \delta_3)$ & $(\alpha_3, \delta_3)$ & $(\alpha_2, \delta_2)$ \\
   & $(\alpha_4, \delta_4)$ & $(\alpha_4, \delta_4)$ & $(\alpha_3, \delta_3)$ \\
\hline
\end{tabular}
\end{center}
\end{table}

Without loss of generality, we use configuration~0 of Table~\ref{tab:ref_point_selection} for all analyses described in Section~\ref{sec:mi_subcomponents}.

\section{Local lens properties for systematically interchanged subcomponent labels}
\label{app:mi_swapping_subcomponents}

\begin{table*}[h!]
 \caption{Local lens properties as obtained using the three subcomponents in images $A$, $C$, and  $D$ with the matching across the images as listed in Table~\ref{tab:systematic_relabelling}.}
\label{tab:swapped_subcomponent_labellings} 
\begin{center}
\begin{tabular}{c|rrr|rrr|rrr}
\hline
Lens prop.	&		&	Conf. 1	&		&		&	Conf. 2	&		&		&	Conf. 3	&		\\
\hline																			
$\mathcal{J}_A$	&	1.00	&	1.00	&	0.00	&	1.00	&	1.00	&	0.00	&	1.00	&	1.00	&	0.00	\\
$f_A$	&	1.00	&	1.00	&	0.00	&	1.00	&	1.00	&	0.00	&	1.00	&	1.00	&	0.00	\\
$g_{A,1}$	&	-0.63	&	-0.63	&	0.05	&	-0.61	&	-0.62	&	0.05	&	-0.51	&	-0.51	&	0.02	\\
$g_{A,2}$	&	0.96	&	0.96	&	0.09	&	1.31	&	1.35	&	0.24	&	0.84	&	0.84	&	0.06	\\
\hline																			
$\mathcal{J}_C$	&	-0.20	&	-0.19	&	0.06	&	0.20	&	0.19	&	0.04	&	-0.20	&	-0.19	&	0.06	\\
$f_C$	&	-0.77	&	-1.51	&	67.44	&	-0.61	&	-0.60	&	109.97	&	-2.29	&	-3.62	&	127.75	\\
$g_{C,1}$	&	0.26	&	-0.40	&	58.19	&	-0.32	&	-0.31	&	2.94	&	-0.75	&	-2.18	&	117.01	\\
$g_{C,2}$	&	0.05	&	0.58	&	48.05	&	-1.72	&	-1.70	&	124.41	&	1.08	&	2.02	&	95.77	\\
\hline																			
$\mathcal{J}_D$	&	-0.17	&	-0.17	&	0.11	&	0.17	&	0.17	&	0.08	&	0.17	&	0.16	&	0.11	\\
$f_D$	&	-0.60	&	-1.28	&	52.54	&	-0.20	&	-0.22	&	1.72	&	2.29	&	-13.26	&	1235.95	\\
$g_{D,1}$	&	0.05	&	-0.14	&	12.08	&	-0.12	&	-0.13	&	0.18	&	-0.37	&	1.53	&	151.29	\\
$g_{D,2}$	&	-0.60	&	-0.25	&	28.05	&	-1.11	&	-1.12	&	0.88	&	0.14	&	-7.68	&	622.89	\\
\hline    
\end{tabular}
\end{center}
\end{table*}

\section{Local lens properties for the four-image configuration $ABCD$}
\label{app:mi_swapping_subcomponents2}

\begin{table*}[h!]
 \caption{Local lens properties, their most likely value, mean, and the $1-\sigma$ uncertainty bound (in columns 1, 2, and 3 of each configuration, respectively), as obtained using the three subcomponents in images $A$, $B$, $C$, and $D$ with the matching across the images as listed in Table~\ref{tab:systematic_relabelling}. Configuration~1 uses interchanged subcomponents $1$ and $3$ in image $B$, Configuration~2 uses a 3 mas uncertainty in the positions of the subcomponents in image $B$, Configuration~3 increases the uncertainty in the subcomponent position to 3 mas for all subcomponent positions of all images.}
\label{tab:swapped_subcomponent_labellings2} 
\begin{center}
\begin{tabular}{c|rrr|rrr|rrr}
\hline
Lens prop.	&		&	Conf. 1	&		&		&	Conf. 2	&		&		&	Conf. 3	&		\\
\hline																			
$\mathcal{J}_A$	&	1.00	&	1.00	&	0.00	&	1.00	&	1.00	&	0.00	&	1.00	&	1.00	&	0.00	\\
$f_A$	&	1.00	&	1.00	&	0.00	&	1.00	&	1.00	&	0.00	&	1.00	&	1.00	&	0.00	\\
$g_{A,1}$	&	-0.63	&	-0.63	&	0.04	&	-0.55	&	-0.56	&	0.04	&	0.03	&	-0.09	&	0.22	\\
$g_{A,2}$	&	1.02	&	1.01	&	0.08	&	0.79	&	0.79	&	0.10	&	-0.91	&	-0.32	&	1.17	\\
\hline																			
$\mathcal{J}_B$	&	0.52	&	0.37	&	1.46	&	0.94	&	1.33	&	4.29	&	-1.45	&	-1.86	&	4.09	\\
$f_B$	&	-0.41	&	-0.23	&	2.85	&	-2.50	&	-3.92	&	93.65	&	-0.72	&	-13.23	&	48.15	\\
$g_{B,1}$	&	0.97	&	0.87	&	1.07	&	-0.33	&	-1.09	&	35.99	&	0.03	&	-1.92	&	8.02	\\
$g_{B,2}$	&	0.45	&	0.41	&	0.63	&	0.66	&	0.37	&	28.82	&	-1.03	&	3.56	&	18.28	\\
\hline																			
$\mathcal{J}_C$	&	0.17	&	0.16	&	0.06	&	0.20	&	0.19	&	0.06	&	0.08	&	0.01	&	1.56	\\
$f_C$	&	-2.86	&	1.50	&	395.80	&	1.30	&	1.53	&	33.65	&	0.05	&	-0.17	&	2.62	\\
$g_{C,1}$	&	0.11	&	-0.38	&	72.33	&	-0.51	&	-0.53	&	8.74	&	0.06	&	-0.01	&	0.68	\\
$g_{C,2}$	&	-4.70	&	0.49	&	465.50	&	0.38	&	0.65	&	39.23	&	-1.00	&	-0.71	&	1.55	\\
\hline																			
$\mathcal{J}_D$	&	-0.15	&	-0.14	&	0.11	&	-0.17	&	-0.16	&	0.11	&	-0.05	&	-0.07	&	1.13	\\
$f_D$	&	-0.41	&	-0.47	&	0.89	&	-330.07	&	-0.60	&	377.50	&	0.02	&	0.31	&	6.31	\\
$g_{D,1}$	&	0.08	&	0.04	&	0.21	&	-83.19	&	-0.01	&	90.84	&	0.07	&	0.16	&	0.99	\\
$g_{D,2}$	&	-0.73	&	-0.69	&	0.53	&	191.67	&	-0.52	&	227.50	&	-1.00	&	-0.92	&	2.14	\\
\hline    
\end{tabular}
\end{center}
\end{table*}

\end{document}